\documentclass[preprint,12pt,a4paper]{elsarticle}
\usepackage[pass]{geometry}
\usepackage[T1]{fontenc}
\usepackage{lmodern}
\usepackage{amsmath,amssymb,booktabs,graphicx,hyperref,xcolor}
\usepackage{listings}
\hypersetup{colorlinks=true,linkcolor=blue,citecolor=blue,urlcolor=blue,
pdftitle={CCGeant4: charge-state-resolved transport and energy-loss fluctuations of relativistic heavy ions in Geant4},
pdfauthor={Andrei Azhakin and Ilya Kudryashov}}
\newcommand{\Eu}{E_{\mathrm{u}}}
\newcommand{\Tcut}{T_{\mathrm{cut}}}
\newcommand{\Tmax}{T_{\max}}
\journal{Computer Physics Communications}
\begin{document}
\begin{frontmatter}
\title{CCGeant4: charge-state-resolved transport and energy-loss fluctuations of relativistic heavy ions in Geant4}
\author[a]{Andrei Azhakin\corref{corresponding}}
\ead{azhakin.am21@physics.msu.ru}
\author[b]{Ilya Kudryashov}
\cortext[corresponding]{Corresponding author.}
\address[a]{Faculty of Physics, M.~V.~Lomonosov Moscow State University, 1 Leninskie Gory, Building 2, Moscow 119991, Russia}
\address[b]{Skobeltsyn Institute of Nuclear Physics, M.~V.~Lomonosov Moscow State University, 1 Leninskie Gory, Building 2, Moscow 119991, Russia}
\date{9 November 2026}

\begin{abstract}
CCGeant4 models energy-loss fluctuations of heavy ions in Geant4, accounting for both charge exchange and relativistic corrections to electron scattering beyond the first Born approximation. We use GLOBAL cross sections for electron capture and loss, ATIMA equilibrium stopping powers and charge-dependent delta-electron spectra calculated from the Dirac equation in a frozen-shell potential. The package also supports tabulated elastic ion--nucleus scattering. The silicon example includes interference between Coulomb and nuclear scattering amplitudes. We compare charge evolution with GLOBAL and energy-loss spectra with MOCADI for matched beams and targets. Experimental comparisons cover published energy-loss measurements and several measurements of energy deposited by xenon ions in silicon. For uranium at 800 MeV/u, the energy-loss variances from our Dirac configuration agree with published screened-ion calculations to within 4\% in Be, Al and Pb. Both calculations lie within the reported experimental uncertainties for all three targets, while calculated bismuth spectra in silver are narrower than measured. Calculations in thin silicon layers show that delta-electron escape can make deposited-energy distributions substantially narrower than primary-ion energy-loss distributions, highlighting the importance of secondary-electron transport for detector-response modelling. The distribution includes the Geant4 source patch, the transport tables and their converters, and a runnable example.

\newpage
\noindent\textbf{Program summary}
\begin{small}
\begin{description}
\item[Program title:] CCGeant4.
\item[CPC Library link to program files:] (to be added by Technical Editor)
\item[Developer's repository link:] \href{https://github.com/andreiazhakin/CCGeant4}{\texttt{andreiazhakin/CCGeant4} on GitHub}.
\item[Licensing provisions:] The MIT licence covers independently authored CCGeant4 code and documentation. Modified Geant4 files and code adapted from Geant4 retain the \href{https://geant4.web.cern.ch/download/license.html}{Geant4 Software License} and original notices.
\item[Programming language:] C++17; Python for data preparation and analysis.
\item[Nature of problem:] Predicting energy deposition by a heavy ion requires both its stopping power and the spectrum of delta electrons that can escape the detector. For high nuclear charges the first Born approximation can misrepresent this spectrum. Electron capture and loss add to the fluctuations because the stopping power depends on the ionic charge. To our knowledge, no existing transport code combines explicit charge-state evolution of fast heavy ions with secondary-particle transport to predict energy-deposition distributions in detector geometries.
\item[Solution method:] Discrete capture and loss events are sampled from GLOBAL cross sections. The supplied stopping tables are normalized to ATIMA equilibrium data. Users can substitute tables normalized to their own equilibrium stopping powers. Screened Dirac cross sections determine explicit delta-electron production and the corresponding subtraction from continuous losses. The Legacy mode uses charge-scaled stopping powers and a Mott correction for delta-electron production. The tabulated elastic scattering model produces discrete nuclear recoils. The supplied silicon tables account for the finite sizes of both nuclei and for interference between the Coulomb and nuclear scattering amplitudes. Benchmarks use GLOBAL and MOCADI. Experimental tests use the energy-loss data of Weick et al.\ and new measurements of energy deposited by xenon ions in silicon.
\item[Additional comments including restrictions:] We recommend explicit charge-state transport for $K/A\gtrsim100$ MeV/u. In Dirac mode, users can set the lower activation energy. Legacy uses a fixed transition from 30 to 40 MeV/u. Dirac tables span $\Eu=30$--2000 MeV. Their use requires an electron production threshold of at least 0.99 keV. Above its upper transition, the Dirac model treats the ion as fully stripped and retains Mott corrections to electron scattering. Gas--solid differences and secondary emission from charge-changing events are not explicitly modelled. The general single-scattering interface includes a silicon reference dataset. Table coverage and numerical conventions are specified in \ref{app:numerics}.
\end{description}

\end{small}
\end{abstract}

\enlargethispage{2\baselineskip}
\end{frontmatter}

\section{Introduction}
Fluctuations of ionization signals limit charge and isotope identification in cosmic-ray instruments. In the silicon-stack spectrometer CRIS, energy-loss fluctuations and multiple scattering limit isotope separation, with a stronger scattering contribution at oblique incidence~\cite{StoneCRIS1998}. The planned TIGERISS instrument combines thin silicon detectors with Cherenkov measurements to distinguish individual elements up to lead~\cite{KrizmanicTIGERISS2025}. Predicting the widths of such detector signals requires a description of both energy-loss fluctuations and secondary-particle transport.

Two heavy ions entering a layer at the same speed and charge can lose different amounts of energy because they capture and lose electrons at different depths. This broadening persists at charge equilibrium, and its magnitude depends on the charge-state populations and the mean distances between charge-changing collisions. When the layer thickness is comparable to the charge-relaxation length, the energy-loss spectrum can also retain a memory of the entrance charge. Weick et al.~\cite{WeickPRL2000,WeickNIM2000,Weick2002,WeickThesis} observed enhanced straggling and asymmetric energy-loss distributions in this regime.

A second effect arises even for a bare ion. At high nuclear charge, $Z_{\mathrm p}\alpha/\beta$ need not be small, and the first Born approximation can give an inaccurate spectrum for energetic delta electrons. For a point nucleus the Lindhard--S\o rensen correction to mean electronic stopping equals the sum of the Mott and relativistic Bloch corrections~\cite{LindhardSorensen1996,ScheidenbergerGeissel1998,Kats2021}. The corresponding prediction for collisional straggling follows from the second moment of the electron energy-transfer spectrum~\cite{Scheidenberger1996,Kats2024}. For a partially stripped ion, bound electrons partially screen the projectile field. We treat them in the frozen-shell approximation used in the calculations of Weick et al.~\cite{WeickNIM2000,Weick2002}.

MOCADI~\cite{Iwasa1997,Iwasa2011,MocadiWeb} simulates the transport of ion beams through ion-optical systems and layers of matter. Its treatment of charge exchange and energy-loss straggling provides a reference for testing the Legacy mode with the same GLOBAL cross sections and ATIMA stopping powers. MOCADI does not explicitly transport secondary electrons and therefore does not predict energy-deposition distributions in a finite detector volume.

CCGeant4 follows the primary ion and its delta electrons through a three-dimensional Geant4 geometry~\cite{Agostinelli2003,Allison2006,Allison2016}. An electron that escapes the sensitive volume carries away part of the primary ion's energy loss. The energy deposited in that volume is therefore a distinct observable.

\section{Physical model}
\subsection{State variables and energy coordinate}
Let $Z_{\mathrm p}$ denote the projectile nuclear charge and $N$ the number of bound electrons. The physical ionic charge is $q=(Z_{\mathrm p}-N)e$. Each track carries its dynamic charge, from which $N$ is obtained in the active charge-exchange interval. The Dirac and charge-exchange models use the energy coordinate
\begin{equation}
 \Eu = m_{\mathrm u}c^2(\gamma-1),\qquad m_{\mathrm u}c^2=931.494\ \mathrm{MeV}.
\end{equation}
This coordinate selects equal speeds for different isotopes and differs slightly from $K/A$, where $A$ is the integer mass number. Experimental beam energies and elastic-table energies are specified in $K/A$.

\subsection{Charge-changing collisions}
The electronic state changes through $N\rightarrow N\pm1$ and $N\rightarrow N\pm2$. For a material composed of atomic species $t$ with number densities $n_t$, the transition rate per unit path length is
\begin{equation}
 R_{NN'}(\Eu)=\sum_t n_t\,\sigma^{(t)}_{NN'}(\Eu).
\end{equation}
The total inverse mean free path is $\lambda_N^{-1}=\sum_{N'\ne N}R_{NN'}$. The final state is selected with probability proportional to its partial rate. The microscopic cross sections come from GLOBAL~\cite{Scheidenberger1998,GlobalWeb}, which is also distributed with LISE++~\cite{TarasovBazin2008}. In its solid-target treatment, GLOBAL uses effective cross sections that approximately account for electron loss from excited states before relaxation and changes in capture probabilities~\cite{Scheidenberger1998}. The finite state space contains $N=0,\ldots,27$.

At fixed energy in a pure element, a stationary distribution $p_N$ obeys
\begin{equation}
 \sum_N p_N Q_{NN'}=0,\qquad \sum_N p_N=1,
\end{equation}
where $Q$ has off-diagonal entries $\sigma_{NN'}$ and diagonal entries equal to minus the outgoing sums. We use these equilibrium fractions to normalize the charge-dependent stopping tables. A transported ion need not be in equilibrium and the initial charge distribution is supplied by the application. In the active charge-exchange interval, a table query for $N>27$ stops a Dirac run. With the supplied GLOBAL tables, Legacy assigns zero charge-exchange cross section to such states.

The charge-exchange process updates the ionic charge without generating secondary particles or assigning changes in shell binding energy to a separate deposition channel. In particular, it does not describe the electrons emitted when the projectile loses a bound electron, including the electron-loss-to-continuum cusp observed experimentally~\cite{Hillenbrand2021}, or photons from capture and subsequent atomic relaxation. Explicit target delta-electron production is treated separately, as described below.

Neglecting the direct energy transfer in capture and loss follows the approximation of Weick et al.~\cite{Weick2002}, who found it small compared with electronic stopping between charge-changing collisions.

As an example, for radiative electron capture (REC), our estimate using GLOBAL cross sections near charge equilibrium supports omitting these photons: the photon-energy second moment per unit path remains below 1\% of that from delta electrons above the cusp energy in Be, Al, Si and Pb at 100--2000 MeV/u, even if all photons are emitted forwards.

Following a nuclear reaction or another selected non-electromagnetic interaction, CCGeant4 resets a surviving projectile to its bare charge, $q=Z_{\mathrm p}e$. Secondary nuclear fragments normally start with the bare charge assigned by Geant4 unless the process that creates them specifies a different charge. These initial charges are model assumptions.

\subsection{Target phase and application range}
The supplied elemental charge-changing cross sections are combined using atomic number densities, without an additional gas--solid model. Excited-state populations and their density-dependent relaxation are not propagated.

Ishikawa et al.~\cite{Ishikawa2023} measured $^{208}$Pb in gaseous and solid targets between 35 and 280 MeV/u and compared the charge-state results with GLOBAL. Differences in equilibrium mean charge and stopping disappeared by 280 MeV/u in those systems. Gas--solid differences in individual electron-loss cross sections were nevertheless observed for nickel at 200 MeV/u~\cite{Ogawa2007}.

The recommended application range is $K/A\gtrsim100$ MeV/u. GLOBAL was adjusted mainly to data above 100 MeV/u, with less accurate applicability down to about 30 MeV/u~\cite{Fettouhi2006}. A configurable lower boundary allows explicit charge-state transport to be restricted to the selected energy range. For energy-loss fluctuations, Dirac corrections are applied only to explicitly generated delta electrons above the production threshold. A suitably low electron production cut is therefore important for reasonable accuracy. Section~\ref{sec:cut-dependence} shows the sensitivity to this choice.

\subsection{Frozen-shell Dirac cross sections}
The hard electron spectrum is obtained by treating a target electron as a free electron scattered by the projectile ion in its rest frame, following the screened-ion approach used for relativistic energy-loss calculations~\cite{WeickNIM2000,Weick2002}. For each directly calculated $(Z_{\mathrm p},N)$ state, a self-consistent, spherically averaged Dirac--Hartree--Fock--Slater (DHFS) density is generated with prescribed shell occupations using the \texttt{dhfs} program of RADIAL~\cite{Salvat2019}. The scattering potential contains the attraction of the point nucleus and the electrostatic Hartree field of this density. Although exchange enters the bound-state DHFS calculation, an exchange interaction with the incident electron is not included in the scattering potential.

The radial Dirac equation in this central potential is solved using a driver based on ELSEPA~\cite{Salvat2005,Elsepa2020}. The resulting phase shifts are summed to obtain the elastic differential cross section. In the ion rest frame, the incident electron has kinetic energy $K_*=(\gamma-1)m_{\mathrm e}c^2$. Its scattering angle $\theta_*$ determines the laboratory energy transfer, in the infinite-projectile-mass approximation, through
\begin{equation}
 T=\Tmax^{\infty}\sin^2\!\frac{\theta_*}{2},\qquad
 \Tmax^{\infty}=2m_{\mathrm e}c^2(\gamma^2-1),\qquad
 \frac{d\sigma_N}{dT}=\frac{4\pi}{\Tmax^{\infty}}\frac{d\sigma_N}{d\Omega_*}.
\end{equation}
The shell density is held fixed during each collision: it screens the nuclear field but does not deform or rearrange in response to the incident electron. This is the frozen-shell approximation. The calculation omits dynamic polarization, excitation and ionization of the projectile shell. Charge-changing collisions are supplied separately by GLOBAL. The point-nucleus assumption is retained in these tables. The finite-size correction in the analytic high-energy branch is a separate approximation.

For each projectile state, the table supplies the differential cross section per target electron,
\begin{equation}
 D_N(\Eu,T)=\frac{d\sigma_N}{dT},
 \qquad [D_N]=\mathrm{cm^2\,MeV^{-1}},
\end{equation}
above $T_0=1$ keV.

\subsection{Continuation in charge state and soft stopping}
\label{sec:charge-continuation}
Direct Dirac spectra are supplied for $N=0,\ldots,N_{\max}$, with $N_{\max}=10$ for $29\le Z_{\mathrm p}\le50$ and $N_{\max}=15$ for $51\le Z_{\mathrm p}\le92$. No further Dirac solutions are calculated for the remaining states. Instead, the last directly calculated spectrum at the same energy is scaled by the squared ionic-charge ratio:
\begin{equation}
 D_N(\Eu,T)=D_{N_{\max}}(\Eu,T)\left(\frac{Z_{\mathrm p}-N}{Z_{\mathrm p}-N_{\max}}\right)^2,
 \qquad N_{\max}<N\le27.
\end{equation}
This continuation retains the shape of the last calculated spectrum.

This approximation targets highly stripped, fast ions: in the pure-element equilibrium calculations used to build the tables, the total population of states with $N>N_{\max}$ is at most $1.26\times10^{-3}$ (0.126\%). For ions with more bound electrons, outer-shell polarization, excitation and configuration dependence also limit the frozen-shell approximation~\cite{Salvat2005}.

For an element $t$, define the hard first moment per target atom by
\begin{equation}
 H_{tN}=Z_t\int_{T_0}^{\Tmax^{\infty}} T D_N(\Eu,T)\,dT .
\end{equation}
For small energy transfers we assume that the collision does not resolve the projectile's electron cloud, so that the target electron sees the net ionic charge $(Z_{\mathrm p}-N)e$. We therefore take the soft stopping to scale as $(Z_{\mathrm p}-N)^2$ and write the full electronic stopping cross section as
\begin{equation}
 \epsilon_{tN}=B_t(\Eu)(Z_{\mathrm p}-N)^2+H_{tN}.
\end{equation}
The coefficient $B_t$ is determined from the stationary GLOBAL distribution and equilibrium ATIMA stopping data~\cite{AtimaWeb}:
\begin{equation}
 B_t=\frac{\epsilon_{t,\mathrm{ATIMA}}-\sum_Np_{tN}H_{tN}}
 {\sum_Np_{tN}(Z_{\mathrm p}-N)^2}.
\end{equation}
Both $B_t$ and $\epsilon_{tN}$ have units of MeV\,cm$^2$ per target atom.

The ATIMA range-derived stopping powers include a small nuclear component, estimated to be below 1\% even at the lower tabulation boundary of 2 MeV/u. We neglect this component when using ATIMA as the electronic stopping reference. Nuclear energy transfer is treated separately by continuous nuclear stopping or, with single scattering, by discrete nuclear recoils.

\subsection{Production threshold and finite projectile mass}
For the pure Dirac component and an electron production threshold $\Tcut$, the continuous stopping subtracts the first moment carried by explicitly generated electrons:
\begin{equation}
 S_{\mathrm{restricted}}=S_{\mathrm{full}}-
 \sum_t n_t Z_t\int_{\Tcut}^{\Tmax(M)}T D_N^{(M)}(\Eu,T)\,dT.
\end{equation}
A projectile of finite mass can transfer slightly less energy to an electron than an infinitely massive projectile at the same speed. We rescale the tabulated electron energies to this lower maximum while preserving the integrated cross section (\ref{app:numerics}). After this rescaling, Geant4's \texttt{G4DeltaAngle} uses the sampled electron energy to generate its laboratory direction, with a uniform azimuth about the ion direction. The electron production threshold must be at least 0.99 keV after conversion of the Geant4 range cut.

\subsection{Transitions and fluctuations}
\label{sec:transitions-fluctuations}
The default Dirac weight $w$ rises from zero to one between 30 and 40 MeV and falls between 1500 and 2000 MeV in $\Eu$, using smoothstep interpolation in logarithmic energy. Users can change these limits. At and below the lower activation energy, explicit charge exchange is disabled. On the low-energy side, the fallback uses tabulated equilibrium stopping powers (ATIMA by default) and a Mott-corrected spectrum for explicit delta electrons down to $\Eu=2$ MeV. Below 2 MeV the model uses Geant4's low-energy stopping treatment.

For candidate delta electrons, the Dirac and fallback contributions to the collision rate are proportional to $w\sigma_{\mathrm D}$ and $(1-w)\sigma_{\mathrm F}$, respectively, where both cross sections are integrated above the production threshold. We select a spectrum in proportion to its contribution and sample the electron energy from it.

Above the Dirac tables, Coulomb corrections remain relevant for high-$Z$ ions because $Z_{\mathrm p}\alpha/\beta$ remains finite as $\beta\rightarrow1$. Measurements at 700--1000 MeV/u show substantial departures from first-Born straggling~\cite{Scheidenberger1996}, as predicted by relativistic scattering theory~\cite{LindhardSorensen1996,Kats2024}. On the high-energy side, both the stopping power and the delta-electron spectrum are calculated analytically, using Geant4's Mott parametrization~\cite{Boschini2011} for electron scattering. The mean stopping includes the Bloch correction and a finite-nuclear-size correction. In the upper transition the projectile form factor can reject an electron only if it was sampled from the fallback spectrum. \ref{app:ls-normalization} compares this high-energy continuation with standard Geant4 and ATIMA.

The track charge is effective at or below the lower boundary, dynamical in the intermediate interval, and fully stripped at or above the upper boundary.

In the full Dirac interval, the default model for continuous energy-loss fluctuations is Geant4's \texttt{G4UniversalFluctuation}. Outside the full Dirac interval, the default mixture selects \texttt{G4IonFluctuations} with probability $1-w$ and Universal with probability $w$. Both models receive the same mean continuous loss for the step, calculated from the blended restricted stopping power (\ref{app:numerics}). In the lower transition, charge exchange is already active, while the probability of using the empirical ion-fluctuation model decreases as the energy rises.

We retain the Universal model's treatment of atomic excitation and ionization below the electron production threshold~\cite{Geant4UniversalFluctuations}. Changing the production cut shifts energy transfers between continuous losses and explicit delta-electron production, so some cut dependence of the energy-loss variance is expected.

\subsection{Legacy configuration}
\label{sec:legacy-model}
The package retains a Legacy configuration for comparison with transport codes using charge-scaled equilibrium stopping. At each table energy, this configuration assigns
\begin{equation}
 S_{\mathrm{Legacy}}(q)=S_{\mathrm{ATIMA}}
       \frac{q^2}{\langle q^2\rangle_{\mathrm{eq}}},
\end{equation}
while following the discrete GLOBAL charge-changing histories. The benchmarks below use this $q^2$ prescription and Universal continuous fluctuations. In these Legacy calculations, the delta-electron spectrum is normalized with the squared instantaneous ionic charge. The Mott coefficients are selected using Geant4's effective charge rounded to the nearest integer~\cite{Lijian1995,Boschini2011}. This charge is usually close to the instantaneous ionic charge in the cases considered. The mean energy carried by these electrons is subtracted from the continuous stopping. The Dirac configuration instead uses the charge-resolved spectra and soft normalization described above. Table~\ref{tab:configurations} summarizes the transport configurations used in this work.

\begin{table}[!htb]
\centering
\caption{Configurations used in this work. The Dirac row describes the full Dirac interval. The transition regions are described in Section~\ref{sec:transitions-fluctuations}.}
\label{tab:configurations}
\small
\setlength{\tabcolsep}{2.8pt}
\renewcommand{\arraystretch}{1.15}
\newcommand{\configcell}[1]{{\raggedright #1\par}}
\begin{tabular}{@{}p{0.17\linewidth}p{0.16\linewidth}p{0.20\linewidth}p{0.19\linewidth}p{0.21\linewidth}@{}}
\toprule
\configcell{Configuration} & \configcell{Charge state} & \configcell{Mean electronic stopping} & \configcell{Delta-electron spectrum} & \configcell{Continuous fluctuations} \\
\midrule
\configcell{Geant4 electromagnetic physics option 4 (\texttt{emopt4})} & \configcell{Effective charge} & \configcell{Standard Geant4} & \configcell{Standard Geant4} & \configcell{Ion\-Fluctuations, with its transition to Universal} \\
\configcell{emopt4 + Universal} & \configcell{Effective charge} & \configcell{Standard Geant4} & \configcell{Standard Geant4} & \configcell{Universal} \\
\configcell{Legacy} & \configcell{Discrete GLOBAL transitions} & \configcell{$q^2$-scaled ATIMA} & \configcell{Mott-corrected, with $q^2$ normalization} & \configcell{Universal} \\
\configcell{Dirac} & \configcell{Discrete GLOBAL transitions} & \configcell{Charge-dependent, normalized to equilibrium ATIMA} & \configcell{Screened Dirac} & \configcell{Universal} \\
\bottomrule
\end{tabular}
\par\smallskip
\begin{minipage}{\linewidth}\small
Standard electron transport and MicroElec are independent choices for transporting the generated secondary electrons.
\end{minipage}
\end{table}

\subsection{Single elastic scattering}
\label{sec:single-scattering}
Applications can select multiple scattering, native Geant4 single scattering or tabulated elastic scattering. The native ion Coulomb model uses a screened Coulomb potential~\cite{Geant4IonCoulomb}. The supplied silicon tables instead include the finite sizes of both nuclei and interference between the Coulomb and nuclear amplitudes. The interface accepts other projectile--target pairs. The silicon dataset covers $Z_{\mathrm p}=26,\ldots,82$ at 207 energies from 10 to 5000 MeV/u.

For each collision the model selects a target isotope in proportion to its number density times the total elastic cross section. It samples the centre-of-mass scattering angle from the tabulated angular distribution and uses two-body kinematics to obtain the outgoing ion and recoil. Isotopes of a given element use the same reference cross sections, but their actual masses enter the kinematics. Recoils above the production threshold are transported, while those below it deposit their energy locally. This discrete transfer replaces continuous nuclear stopping for ions assigned to single scattering. Outside the table coverage the model uses native Geant4 screened Coulomb scattering.

The construction and use of the silicon reference tables are described in \ref{app:elastic}.

\section{Implementation}
The converter combines Dirac spectra, GLOBAL transition cross sections and stopping powers obtained by differentiating ATIMA 1.41 range splines~\cite{AtimaWeb,ScheidenbergerGeissel1998,LindhardSorensen1996}. \ref{app:numerics} describes the interpolation and the checks on the tabulated quantities. File formats and conversion commands are documented with the software.

The patch adds charge exchange and charge reset, and modifies electronic stopping and delta-electron production to account for the instantaneous ionic charge. The high-energy stopping and delta-electron models are also revised, and an optional tabulated model for elastic nuclear scattering is added. Applications select tables and transition boundaries before initialization through \texttt{G4EmStandardPhysics\_\allowbreak option4\_\allowbreak CurrentCharge}, replacing the electromagnetic constructor of a modular physics list. The patch requires rebuilding Geant4. The README provides installation instructions and a runnable example. Threads share the physics tables. The implementation uses Geant4 11.4.0, and patch application to 11.4.3 has been checked.

Charge-changing collisions occur at the end of a transport step. Continuous energy loss over that step uses the charge state before the collision. On the next step, stopping, charge-changing rates and delta-electron production are evaluated using the updated track charge.

For Cu, Xe and U, we checked that sampled delta-electron energies lay between the production threshold and the kinematic maximum and compared their mean with that of the tabulated spectrum. We also checked the balance of continuous and explicit losses, event-energy accounting and agreement between single- and multithreaded runs. The example includes reference transport results.

OpenAI Codex with GPT 6 Astra (OpenAI) assisted code development and review and preparation of the analysis and plotting scripts. The figures were generated from simulation outputs, reference calculations and experimental data using these scripts.

At a 25~$\mu$m production cut, representative uranium benchmarks required 1.7--3.7 times the single-thread transport time of \texttt{emopt4}, with a 6.3--6.8-fold speed-up on eight threads. The run conditions, timings and memory use are given in \ref{sec:performance}.

\section{Comparison with GLOBAL and MOCADI}
\label{sec:legacy-cross-code}
GLOBAL provides a reference for the charge fractions predicted by both modes. For the energy-loss comparison with MOCADI we use Legacy, which shares its $q^2$ stopping prescription (Section~\ref{sec:legacy-model}).

\subsection{Charge evolution in silicon}
We compare the predicted charge fractions with standalone GLOBAL (May 2017) for
$^{209}\mathrm{Bi}$ at $380.416\,\mathrm{MeV}$ per nucleon, incident on
silicon in the pure states $q=83$ and $q=81$. The Dirac and Legacy modes use the same GLOBAL transition cross sections but different charge-dependent stopping powers, so their exit fractions need not be identical as the beam slows down. GLOBAL uses its native range--energy relation, while the two Geant4 modes use stopping powers normalized to ATIMA.

\begin{figure}[tbp]
\centering
\includegraphics[width=\linewidth]{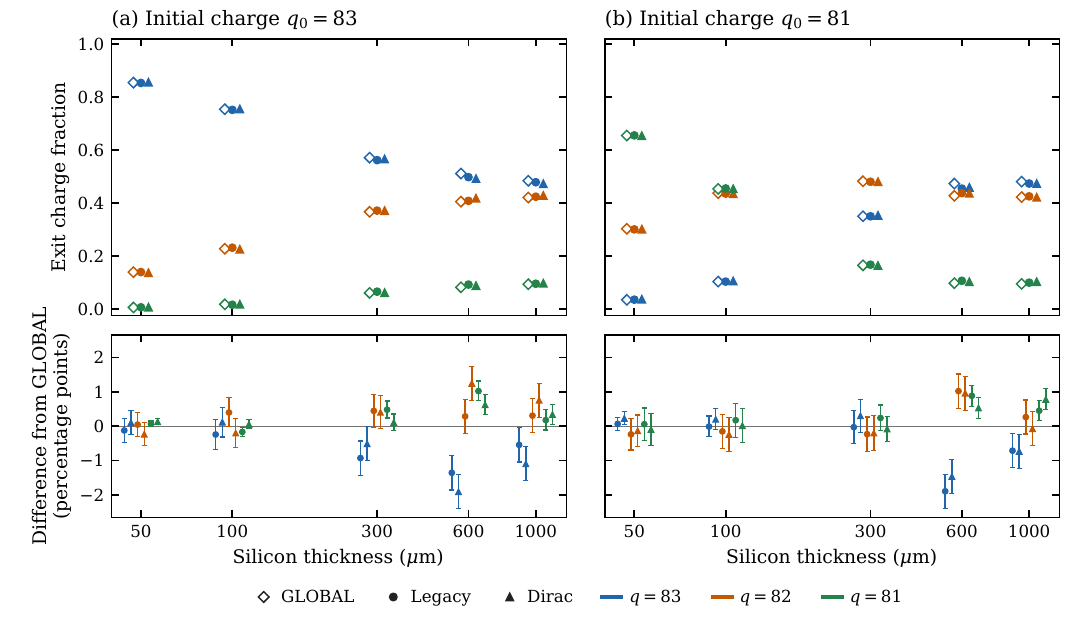}
\caption{Exit charge fractions of $^{209}\mathrm{Bi}$ in silicon at $380.416$ MeV/u for initial charges (a) $83+$ and (b) $81+$. Top: fractions from GLOBAL, Legacy and Dirac for three outgoing charges. Bottom: CCGeant4 minus GLOBAL in percentage points. Bars show 68\% confidence intervals for the CCGeant4 fractions. Each CCGeant4 sample contains $10\,000$ ions. Symbols at the same thickness are offset horizontally for clarity.}
\label{fig:legacy-global}
\end{figure}

After $300\,\mu\mathrm{m}$ of silicon the entrance charge still matters (Fig.~\ref{fig:legacy-global}): Dirac gives bare-ion fractions of $56.6\%$ and $35.4\%$ for initially bare and heliumlike bismuth, respectively. Legacy gives $56.2\%$ and $35.0\%$. By $1000\,\mu\mathrm{m}$ both incident beams give a bare-ion fraction near $47\%$ in Dirac. Here the charge distribution approaches equilibrium over a macroscopic distance, of order a millimetre.

Over the five thicknesses the largest difference between Dirac and Legacy is 0.96 percentage points: at $600\,\mu\mathrm{m}$, $41.80\%$ of initially bare ions exit as $82+$ in Dirac, compared with $40.84\%$ in Legacy. The largest deviations from GLOBAL are 1.91 percentage points for Dirac and 1.89 for Legacy.

\subsection{Energy-loss straggling}
We compare primary-ion
energy-loss spectra from MOCADI 4.2g~\cite{Iwasa1997,Iwasa2011}, using GLOBAL and ATIMA 1.41,
with the Legacy calculation for bismuth
at $380.416\,\mathrm{MeV}$ per nucleon in 150 and $300\,\mu\mathrm{m}$
of silicon, and for xenon and bismuth at $1000\,\mathrm{MeV}$ per nucleon
in $300\,\mu\mathrm{m}$ of silicon.
All MOCADI calculations use $F_{\mathrm{chstr}}=1$, selecting the built-in
GLOBAL cross sections and $q^2$ scaling of the charge-dependent stopping power.

For bismuth at $380.416\,\mathrm{MeV}$ per nucleon, both codes use the same
initial equilibrium mixture, with approximately $50.0\%$ bare,
$41.2\%$ hydrogenlike and $8.7\%$ heliumlike ions. The higher-energy
benchmarks start with bare ions. All beams enter the silicon normally in
vacuum. The CCGeant4 benchmarks use standard electron transport and
production cuts of $25\,\mu\mathrm{m}$ for the lower-energy bismuth series
and $15\,\mu\mathrm{m}$ for the $1000\,\mathrm{MeV}$ per nucleon series.
There are $10^4$ events per code and thickness in the first series and
$10^5$ per code and ion in the second.

\begin{figure}[tbp]
\centering
\includegraphics[width=\linewidth]{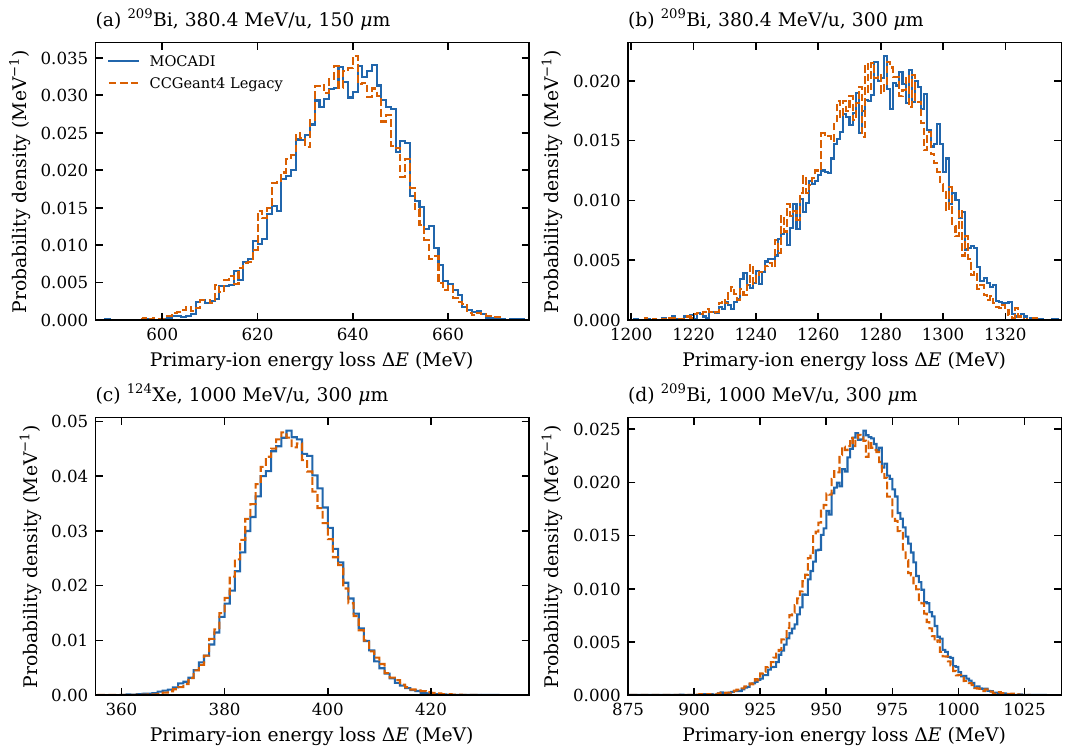}
\caption{Primary-ion energy loss in silicon calculated with MOCADI and
CCGeant4 Legacy with explicit charge exchange and Mott-corrected electron production.
The spectra are normalised to unit area and plotted on the absolute energy
scale without shifts.}
\label{fig:legacy-mocadi}
\end{figure}

Figure~\ref{fig:legacy-mocadi} compares the energy-loss spectra.
Table~\ref{tab:legacy-mocadi} reports their full standard deviations and the
deposited-energy widths from CCGeant4.
Widths are sample standard deviations of the complete distributions, without tail removal or Gaussian fitting. Bootstrap intervals describe Monte Carlo statistical uncertainty.
The central values of the full primary-ion energy-loss standard deviations
differ by less than $2\%$ in all four cases. With GLOBAL cross sections
and ATIMA stopping data common to both codes, this agreement tests their transport
implementations. Mean losses in
MOCADI are higher by $0.04$--$0.19\%$. Independent tabulation grids,
interpolation and slowing-down conventions can contribute to residual
differences at this scale.

\begin{table}[tbp]
\centering\small
\caption{Full standard deviations of energy loss in MOCADI
($\sigma_{\mathrm{M}}$) and the Legacy mode ($\sigma_{\mathrm{G}}$),
and of energy deposited in silicon in CCGeant4 ($\sigma_{\mathrm{G,dep}}$).
Ratio uncertainties are rounded 68\% bootstrap intervals.}
\label{tab:legacy-mocadi}
\begin{tabular}{lrrrrrr}
\toprule
Ion & $\varepsilon$ & $t_{\mathrm{Si}}$ & $\sigma_{\mathrm{M}}$ &
$\sigma_{\mathrm{G}}$ & $\sigma_{\mathrm{M}}/\sigma_{\mathrm{G}}$ &
$\sigma_{\mathrm{G,dep}}$ \\
 & (MeV/u) & ($\mu$m) & (MeV) & (MeV) & & (MeV) \\
\midrule
Bi & 380.416 & 150 & 11.644 & 11.679 & $0.997\pm0.009$ & 8.144 \\
Bi & 380.416 & 300 & 19.254 & 18.917 & $1.018\pm0.010$ & 15.031 \\
Xe & 1000 & 300 & 8.409 & 8.452 & $0.995\pm0.003$ & 3.003 \\
Bi & 1000 & 300 & 16.587 & 16.573 & $1.001\pm0.003$ & 5.726 \\
\bottomrule
\end{tabular}
\end{table}

The deposited-energy columns in Table~\ref{tab:legacy-mocadi} illustrate secondary-electron escape from the silicon layer. \ref{app:legacy-deposition} shows the corresponding Legacy spectra and their entrance-charge decomposition.

\section{Energy-loss validation}
\label{sec:energy-loss-validation}
\subsection{Data and event selection}
We compare 48 Dirac and 41 Legacy simulations with the results of Weick et al.~\cite{WeickPRL2000,WeickNIM2000,Weick2002,WeickThesis}. Each simulation uses 14\,900 events. Common cases use identical beams, targets, areal densities and charge selections. Monoenergetic ions enter normally. Multiple scattering is included and hadronic reactions are disabled. A production range cut of $10~\mu$m is used for Dirac, Legacy and the standard Geant4 controls. The energy loss is $\Delta E=T_{\mathrm{in}}-T_{\mathrm{out}}$. We select primary ions leaving the downstream face with unchanged nuclear identity $(A,Z)$, without an angular cut or outlier rejection. The exit-charge selection follows the criterion of the corresponding Weick study~\cite{WeickPRL2000,Weick2002,WeickThesis}.

``Weick Point'' denotes the published point-charge calculation, which uses charge-dependent Lindhard--S\o rensen stopping and the bare nuclear charge for collision straggling~\cite{WeickThesis}.

For Bi, target areal densities were obtained by integrating the inverse equilibrium stopping power between the reported entrance and mean exit energies, then refined in short simulations to reproduce the mean exit energy. The same densities were used for all compared configurations. The calculations use pure-element targets.

We digitized the reference values and experimental error bars from the figures of Weick et al. The finite size of the plotted symbols limits the reading precision, especially for small values on a wide vertical scale. To estimate the Monte Carlo statistical errors, we repeatedly drew new samples with replacement from the simulated events and used the standard deviation of the resulting estimates.

\subsection{Uranium energy-loss variances}
\begin{figure}[tbp]
\centering
\includegraphics[width=\linewidth]{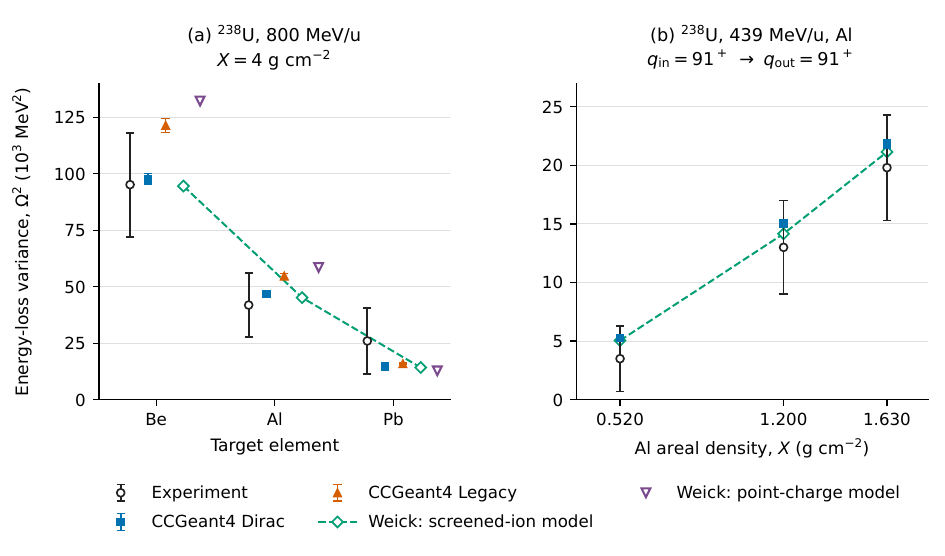}
\caption{Uranium energy-loss variances. The left panel shows $^{238}$U at 800 MeV/u in 4 g/cm$^2$ of Be, Al and Pb. Measurements and Weick Point values are from Fig.~4.10 of Ref.~\cite{WeickThesis}. The right panel shows $^{238}$U at 439 MeV/u in Al with $91+\rightarrow91+$ selection, with measurements from Fig.~2 of Ref.~\cite{Weick2002}. Screened-model predictions are from Ref.~\cite{Weick2002}. Experimental bars reproduce the published bars. Dirac and Legacy bars show Monte Carlo statistical errors. Charge selections are given in Table~\ref{tab:weick-u}.}
\label{fig:weick-u}
\end{figure}

Figure~\ref{fig:weick-u} and Table~\ref{tab:weick-u} show six uranium comparisons. At 800 MeV/u and a target areal density of 4 g/cm$^2$, both configurations reproduce the measured decrease in variance from Be to Al to Pb. At 439 MeV/u in Al, the variance increases with target thickness for the $91+\rightarrow91+$ selection.

At 800 MeV/u the Dirac variances differ from the screened-ion calculations of Weick et al.~\cite{Weick2002} by about 3--4\% in Be, Al and Pb. Both sets of predictions lie within the experimental error bars in all three targets. The relative experimental uncertainty is largest in Pb, where the measured variance is approximately $(26.0\pm14.5)\times10^3$~MeV$^2$. At 439 MeV/u the two calculations agree to within about 6\% at all three Al thicknesses.

\begin{table}[tbp]
\centering
\small
\setlength{\tabcolsep}{2.8pt}
\caption{Uranium variance comparisons. $E=K/A$ is in MeV/u; $X$ is in g/cm$^2$; all variances are in $10^3$ MeV$^2$. Calculation errors are Monte Carlo statistical. Experimental uncertainties are reproduced from the sources. Charges are in units of $e$.}
\label{tab:weick-u}
\begin{tabular}{rrcrrrr}
\toprule
$E$ & $X$ & Target & $q_{\rm in}\!\to q_{\rm out}$ & Dirac & Legacy & Experiment \\
\midrule
800 & 4.000 & Be & $91\to91$ & $97.710\pm2.270$ & $121.392\pm3.039$ & $95.1\pm23.0$ \\
800 & 4.000 & Al & $92\to91$ & $46.710\pm1.020$ & $54.696\pm1.275$ & $41.9\pm14.2$ \\
800 & 4.000 & Pb & $92\to92$ & $14.747\pm0.316$ & $15.943\pm0.343$ & $26.0\pm14.5$ \\
439 & 0.520 & Al & $91\to91$ & $5.271\pm0.083$ & --- & $3.5\pm2.8$ \\
439 & 1.200 & Al & $91\to91$ & $15.048\pm0.240$ & --- & $13.0\pm4.0$ \\
439 & 1.630 & Al & $91\to91$ & $21.838\pm0.392$ & --- & $19.8\pm4.5$ \\
\bottomrule
\end{tabular}
\end{table}

\subsection{Bismuth energy-loss variances}
\begin{figure}[tbp]
\centering
\includegraphics[width=\linewidth]{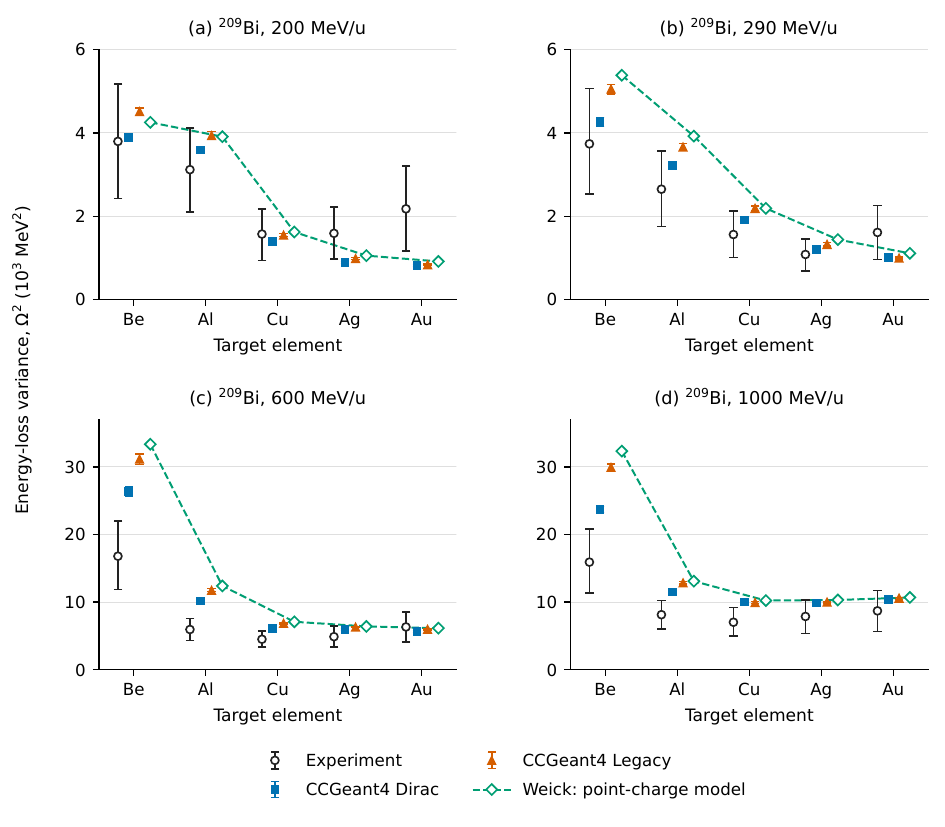}
\caption{Energy-loss variances for $^{209}$Bi at four incident energies in five target elements. Measurements are from Refs.~\cite{WeickPRL2000,WeickThesis}. All point-charge predictions are from Fig.~4.9 of Ref.~\cite{WeickThesis}. Our Dirac and Legacy calculations use the same target areal densities and nominal charge selections. Dirac and Legacy bars show Monte Carlo statistical errors.}
\label{fig:weick-bi}
\end{figure}

Figure~\ref{fig:weick-bi} shows 20 bismuth variance cases at 200, 290, 600 and 1000 MeV/u in Be, Al, Cu, Ag and Au. The 200--600 MeV/u points were digitized from Fig.~3 of Ref.~\cite{WeickPRL2000}, and the 1000 MeV/u points from Fig.~4.9 of Ref.~\cite{WeickThesis}.

The ratio of the Legacy variance to the Weick Point variance is close to unity across the 20 bismuth cases. The Dirac variances lie within the experimental error bars in 12 cases, including all five targets at 290 MeV/u. At 600 MeV/u in Al, the Legacy and Dirac variances are $11\,747\pm212$ and $10\,181\pm186$ MeV$^2$, respectively, compared with $5\,950\pm1\,630$ MeV$^2$ from the digitized measurement. At 200 MeV/u in Au, the corresponding values are $838\pm17$, $814\pm18$ and $2\,180\pm1\,020$ MeV$^2$.

For $^{209}$Bi in Be at 200 MeV/u, the measured energy-loss variance is $7.8\pm2.9$ times the Lindhard--S\o rensen collisional variance~\cite{WeickNIM2000}. The corresponding ratios are 7.2 in the screened-ion calculation shown in Fig.~47 of Ref.~\cite{GeisselWeick2002}, 7.5 in Dirac and 9.0 in Legacy (\ref{app:bi-normalized}, Fig.~\ref{fig:weick-bi-normalized}).

For $^{209}$Bi at 1000~MeV/u in Cu, Ag and Au, we compare Dirac and Legacy with Geant4~11.4.0 \texttt{emopt4} at the same target thicknesses, including all transmitted ions regardless of their exit charge. In all three targets, Dirac and Legacy give energy-loss variances 2.6--2.8 times those from \texttt{emopt4} (Table~\ref{tab:weick-vanilla}).

\begin{table}[tbp]
\centering\small
\caption{Energy-loss variances for $^{209}$Bi at 1000 MeV/u, in $10^3$ MeV$^2$. Each value uses all 14\,900 ions that crossed the target, regardless of their exit charge. Errors are Monte Carlo statistical uncertainties.}
\label{tab:weick-vanilla}
\begin{tabular}{lrrr}
\toprule
Target & Dirac & Legacy & Geant4 \texttt{emopt4} \\
\midrule
Cu & $9.893\pm0.122$ & $10.002\pm0.113$ & $3.626\pm0.038$ \\
Ag & $9.842\pm0.117$ & $9.975\pm0.119$ & $3.720\pm0.046$ \\
Au & $10.361\pm0.117$ & $10.641\pm0.120$ & $3.935\pm0.042$ \\
\bottomrule
\end{tabular}
\end{table}

\subsection{Energy-loss distributions and spectral widths}
\begin{figure}[!tp]
\centering
\includegraphics[width=\linewidth]{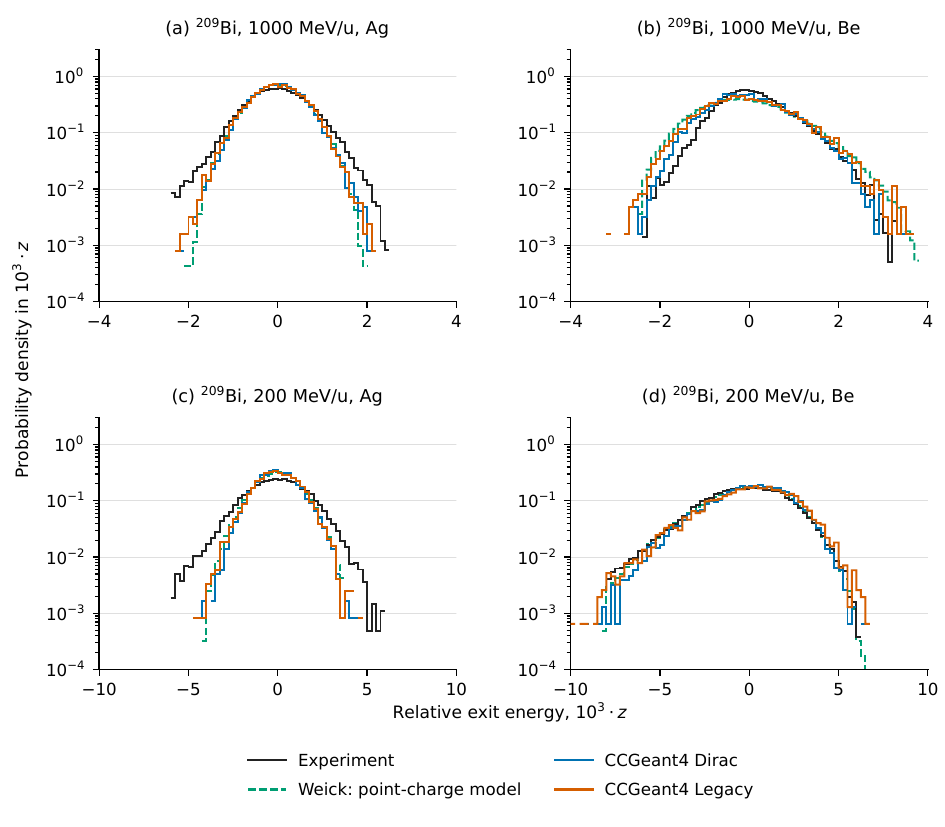}
\caption{Centred relative outgoing-energy spectra for $^{209}$Bi in Ag and Be at 1000 and 200 MeV/u. The experimental and Weick Point histograms are digitized from Fig.~4.13 of Ref.~\cite{WeickThesis}. Each displayed distribution is normalized to unit area over the published interval in $10^3\,\cdot\,z$.}
\label{fig:weick-spectra}

\medskip
\begingroup
\makeatletter\def\@captype{table}\makeatother
\small
\caption{Ratios of calculated to measured central 68\% half-widths for the four $^{209}$Bi spectra in Fig.~\ref{fig:weick-spectra}, in panel order (a)--(d). $n_{\rm D}$ and $n_{\rm L}$ are the selected event counts for Dirac and Legacy. Errors are Monte Carlo statistical uncertainties.}
\label{tab:weick-spectra}
\setlength{\tabcolsep}{4pt}
\begin{tabular}{rcrrrr}
\toprule
$K/A$ (MeV/u) & Target & $q_{\rm in}\!\to q_{\rm out}$ & $n_{\rm D}/n_{\rm L}$ & Dirac & Legacy \\
\midrule
1000 & Ag & $83\to83$ & 12477/12405 & $0.827\pm0.007$ & $0.864\pm0.008$ \\
1000 & Be & $83\to83$ & 6193/6209 & $1.144\pm0.016$ & $1.299\pm0.017$ \\
200 & Ag & $81\to81$ & 4800/4838 & $0.702\pm0.010$ & $0.727\pm0.010$ \\
200 & Be & $81\to81$ & 6154/6174 & $0.957\pm0.014$ & $1.021\pm0.011$ \\
\bottomrule
\end{tabular}
\endgroup
\end{figure}

The four spectral comparisons use the visible published $^{209}$Bi histograms in Fig.~4.13 of Ref.~\cite{WeickThesis}. Define the centred relative outgoing-energy coordinate and its central 68\% half-width by
\begin{equation}
 z=\frac{T_{\mathrm{out}}-\langle T_{\mathrm{out}}\rangle}{\langle T_{\mathrm{out}}\rangle},
 \qquad W_{68}=\frac{Q_{0.84}(z)-Q_{0.16}(z)}{2}.
\end{equation}
Figure~\ref{fig:weick-spectra} compares the centred spectra, and Table~\ref{tab:weick-spectra} gives the ratios of calculated to measured $W_{68}$.

Both configurations reproduce the narrowing of the Be and Ag spectra between 200 and 1000 MeV/u. In Be, the Dirac half-width differs from the experimental value by about 4\% at 200 MeV/u and 14\% at 1000 MeV/u. In Ag, the calculated half-widths are 70--73\% of the measured value at 200 MeV/u and 83--86\% at 1000 MeV/u.

\subsection{Asymmetry}
\begin{figure}[!htbp]
\centering
\includegraphics[width=\linewidth]{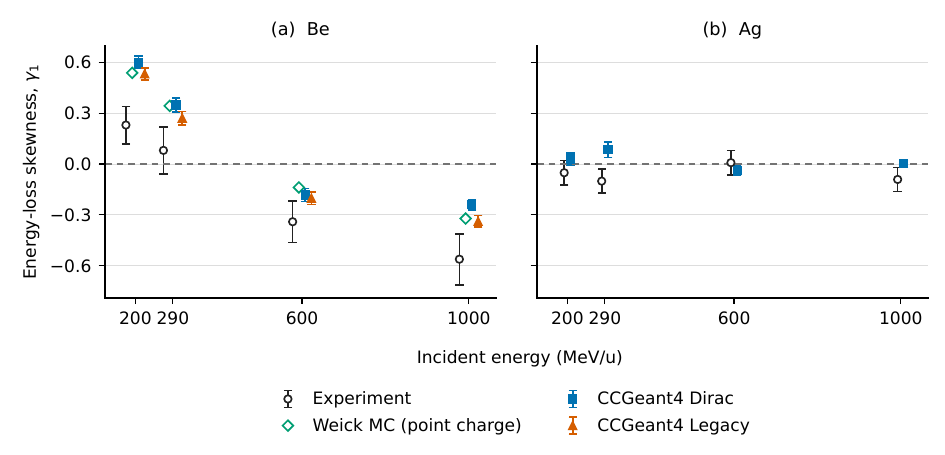}
\caption{Skewness of the energy lost by $^{209}$Bi in Be and Ag at a mean loss of about 20\% of the incident energy. The measurements and the Weick Monte Carlo points for Be are from Fig.~4 of Ref.~\cite{WeickPRL2000}. CCGeant4 bars show statistical uncertainties. Symbols at the same energy are offset for clarity.}
\label{fig:weick-bi-skewness}
\end{figure}

Figure~\ref{fig:weick-bi-skewness} shows how the shape of the $^{209}$Bi energy-loss distribution depends on the target. In Be the measured skewness changes from positive at 200 MeV/u to negative at 1000 MeV/u. Both CCGeant4 configurations reproduce this reversal, with Legacy closer to the measured central values at all four energies. At 200 MeV/u the skewness measurement corresponds to a mean loss of 20\% of the incident energy, compared with 28\% for the Be spectrum in Fig.~\ref{fig:weick-spectra}(d). For the 20\% case, the measured skewness is $0.23\pm0.11$, compared with 0.54 in the Weick simulation, $0.60\pm0.04$ in Dirac and $0.53\pm0.04$ in Legacy~\cite{WeickPRL2000}. In Ag the skewness remains small. The largest difference occurs at 290 MeV/u, where Weick et al.~\cite{WeickPRL2000} measure $-0.101\pm0.070$ and Dirac gives $+0.085\pm0.046$.

\section{Comparison with silicon-detector measurements}
\label{sec:xe-deposition}
For each $^{124}$Xe ion we sum the energy deposited in silicon by the ion and its secondary particles and compare the resulting spectrum with the detector signals measured in Ref.~\cite{AzhakinXeDraft}. We characterize each peak by its relative Gaussian-core width, $100\sigma_{\mathrm{core}}/\mu_{\mathrm{core}}$. The Gaussian core is fitted in an iteratively adjusted window near $\mu_{\mathrm{core}}\pm2\sigma_{\mathrm{core}}$, following Ref.~\cite{AzhakinXeDraft}.

Each simulated sample contains $10^4$ monoenergetic ions striking the centre of the silicon plate at normal incidence. Both detector comparisons use $32\times32$~mm$^2$ silicon plates and GLOBAL equilibrium entrance-charge fractions for Dirac and Legacy. Nuclear reactions are disabled, and no charge or angular selection is applied to the simulated events. In silicon, the 25~$\mu$m production cut corresponds to an electron threshold of about 54~keV. We transport electrons with MicroElec from this threshold to 9~MeV and deposit their remaining energy locally below the threshold~\cite{Valentin2012,Raine2014,AzhakinXeDraft}.

\begin{figure}[tbp]
\centering
\includegraphics[width=\linewidth]{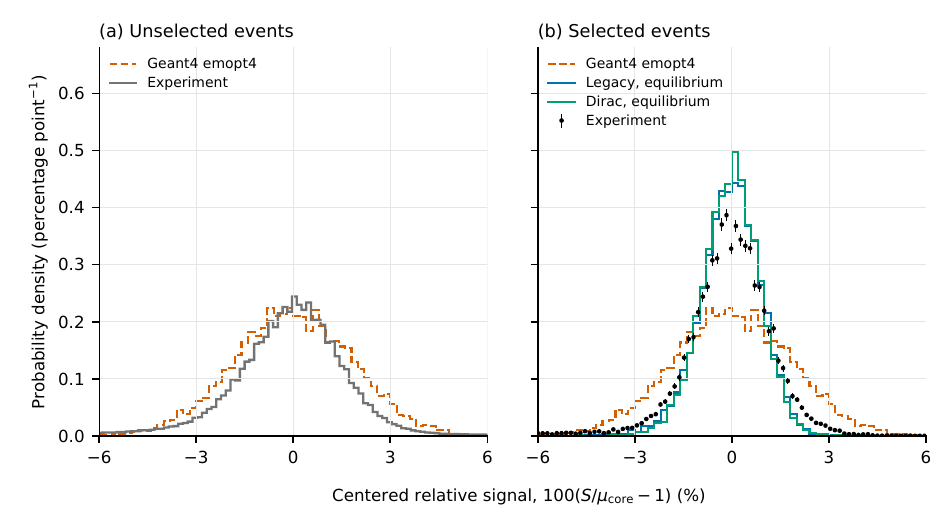}
\caption{Xenon signals in 300~$\mu$m silicon near 267 MeV/u, expressed relative to each spectrum's fitted peak. Panel (a) compares the unselected data with \texttt{emopt4}. Panel (b) compares the selected data with \texttt{emopt4}, Dirac and Legacy. Experimental spectra are from Ref.~\cite{AzhakinXeDraft}.}
\label{fig:xe-dps300}
\end{figure}

In the 300~$\mu$m DPS-NICA detector near 267 MeV/u, restricting tracks to the centre of the pad suppresses edge effects and reduces broadening from variations in silicon thickness. The angular selection excludes tracks near the channeling directions of the pad. After event selection, the measured width falls from 1.41\% to $(1.060\pm0.008)\%$~\cite{AzhakinXeDraft} (Fig.~\ref{fig:xe-dps300}). The remaining broadening from the beam and detector response cannot be fully separated from the intrinsic energy-deposition fluctuations in these data. Dirac and Legacy give $(0.847\pm0.010)\%$ and $(0.885\pm0.010)\%$, respectively. Standard Geant4 \texttt{emopt4} gives $(1.84\pm0.02)\%$, a broader peak than either experimental sample.

\begin{figure}[tbp]
\centering
\includegraphics[width=\linewidth]{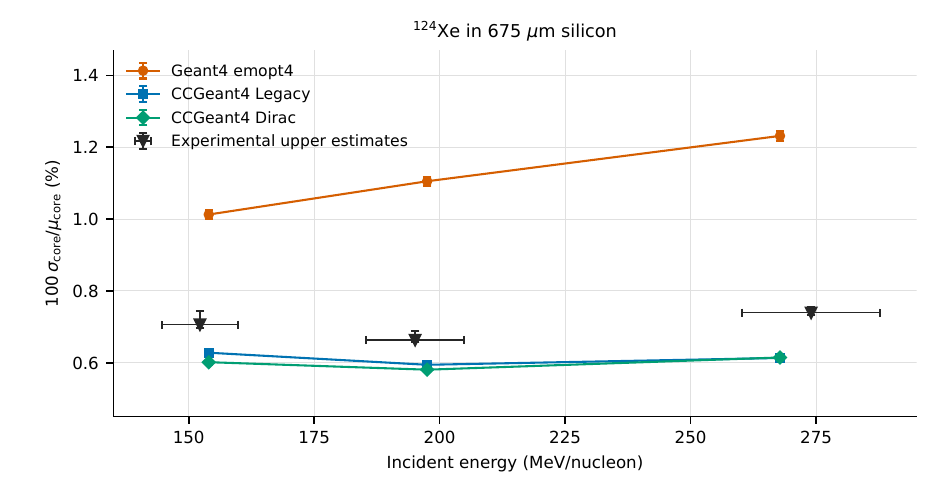}
\caption{Relative peak widths for $^{124}$Xe in 675~$\mu$m silicon. The STIKL vertical bars combine statistical errors and the upward selection uncertainty in quadrature. Horizontal bars show the $\pm5\%$ experimental energy-scale uncertainty~\cite{AzhakinXeDraft}. Uncertainties on the calculated widths are statistical.}
\label{fig:xe-legacy-experiment}
\end{figure}

In the 675~$\mu$m STIKL detector, the selected experimental peaks have widths of $(0.706\pm0.008)\%$, $(0.663\pm0.006)\%$ and $(0.739\pm0.005)\%$ at about 152, 195 and 274 MeV/u, respectively~\cite{AzhakinXeDraft}. Figure~\ref{fig:xe-legacy-experiment} compares them with simulations at 154, 197.5 and 267.8 MeV/u. Dirac and Legacy give similar widths, 0.58--0.63\% across the three energies. The \texttt{emopt4} widths range from 1.01\% to 1.23\% and exceed the measured widths at every energy.

\enlargethispage{\baselineskip}
\looseness=-1
The measured widths provide experimental upper estimates of the intrinsic energy-deposition fluctuations, and the CCGeant4 results are compatible with these estimates. A quantitative test of the calculated widths would require the remaining beam and detector broadening to be determined separately.

\section{Energy dependence of energy-loss and deposition fluctuations}
\label{sec:edep-energy-scan}
\subsection{Comparison with standard Geant4}
\label{sec:standard-em-comparison}
We compare the widths of the energy-loss and deposited-energy distributions across the energies where standard Geant4 changes its treatment of ion fluctuations. Below a projectile-dependent threshold, \texttt{G4IonFluctuations} uses a variance model that includes the empirical charge-fluctuation contribution of Yang et al.~\cite{Yang1991}. Above the threshold, it switches to \texttt{G4UniversalFluctuation}~\cite{Geant4IonFluctSource}. The same ion-fluctuation model, including its transition to Universal fluctuations, is also used by the other standard Geant4 electromagnetic options~\cite{Geant4EmFluctuationSelection}.

Monoenergetic $^{124}$Xe, $^{209}$Bi and $^{63}$Cu beams enter a 300~$\mu$m layer of silicon normally in vacuum. The transverse dimensions are $1\times1$ m$^2$, and the production range cut is 25~$\mu$m. The simulations include nuclear stopping and standard electron transport. Hadronic reactions are disabled. CCGeant4 samples the incident charge from the GLOBAL equilibrium distribution in silicon. Standard Geant4 11.4.0 \texttt{emopt4} starts with a bare ion and uses its native effective-charge prescription. The incident energies span 40--1000~MeV/u, with 10\,000 events per point. At the low-energy end, slowing in silicon brings ions into the $\Eu=30$--40 MeV blending interval, where the fluctuation curves approach those of standard Geant4 as its continuous-fluctuation model is gradually restored. The approximate incident energies at which the mean exit energy reaches $\Eu=40$ MeV are 54, 58 and 49~MeV/u for Xe, Bi and Cu, respectively. These boundaries are marked on the plots.

To suppress rare large-angle and backscattering artifacts from Urban multiple scattering, all widths in this section use the 99\% of transmitted ions with the smallest exit angles.

\begin{figure}[tbp]
\centering
\includegraphics[width=\linewidth]{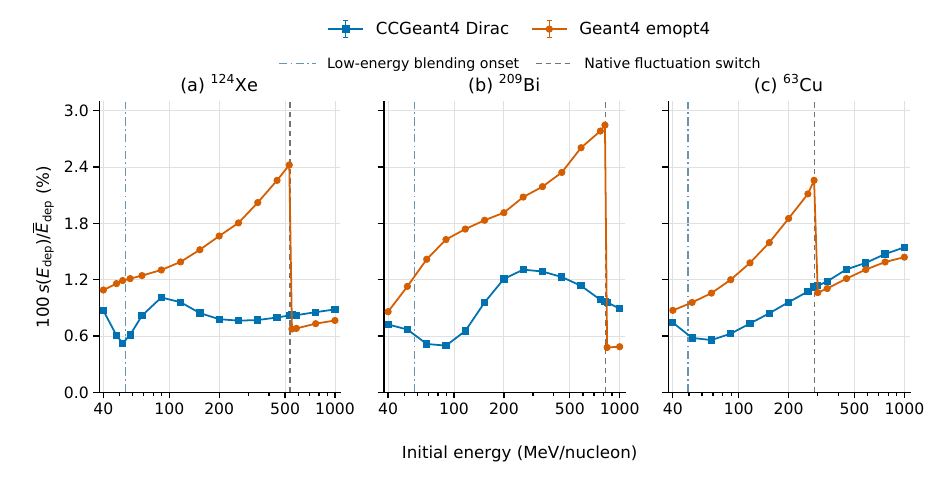}
\caption{Relative deposited-energy widths in 300~$\mu$m of silicon for xenon, bismuth and copper. Grey dashed lines mark the energies at which \texttt{G4IonFluctuations} switches to \texttt{G4UniversalFluctuation}. Blue dash-dotted lines mark the approximate onset of low-energy blending as the ions slow in silicon. Connecting lines guide the eye.}
\label{fig:si-edep}
\end{figure}

Figure~\ref{fig:si-edep} reports the relative width
\begin{equation}
 R_{\mathrm{dep}}=100\,\frac{s(E_{\mathrm{dep}})}{\overline E_{\mathrm{dep}}}\quad(\%),
\end{equation}
where $s$ is the standard deviation of the selected sample.

The standard-Geant4 curves show sharp changes near the switch energies of 535.6, 823.7 and 287.5 MeV/u for Xe, Bi and Cu, respectively, corresponding to the approximately $10Z_{\mathrm p}$ MeV/u threshold in \texttt{G4IonFluctuations}~\cite{Geant4IonFluctSource}. For Xe, $R_{\mathrm{dep}}$ changes from $2.421\pm0.017\%$ at 530 MeV/u to $0.674\pm0.005\%$ at 550 MeV/u. This behaviour is consistent with the sensitivity reported for the Geant4 ion-fluctuation model~\cite{Geant4FluctuationsReport2023}.

At 1000 MeV/u, CCGeant4 predicts broader deposited-energy distributions for all three ions. The respective $R_{\mathrm{dep}}$ values for Xe, Bi and Cu are $0.882\%$, $0.890\%$ and $1.544\%$, compared with $0.764\%$, $0.485\%$ and $1.439\%$ from standard Geant4.

\begin{figure}[tbp]
\centering
\includegraphics[width=\linewidth]{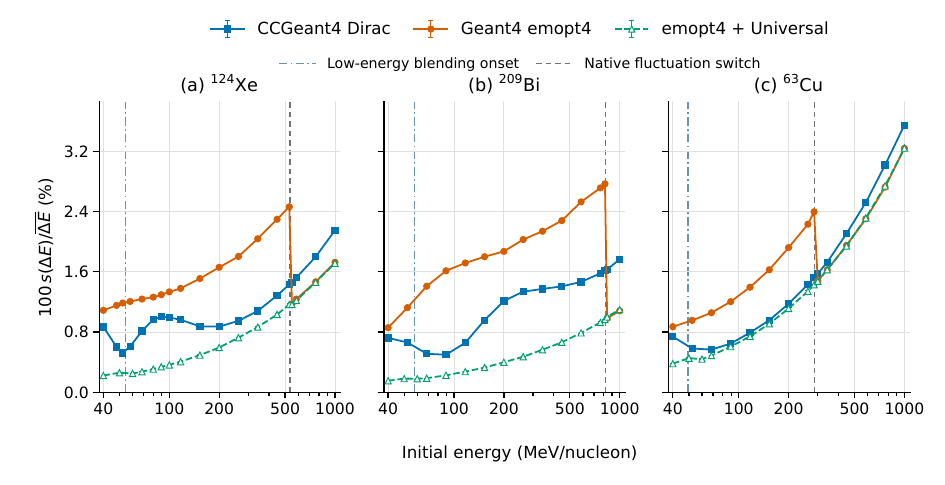}
\caption{Relative widths $100s(\Delta E)/\overline{\Delta E}$ of the primary-ion energy loss in 300~$\mu$m of silicon for CCGeant4, standard \texttt{emopt4} and \texttt{emopt4} with Universal fluctuations. Grey dashed lines mark the native ion-fluctuation thresholds. Blue dash-dotted lines mark the low-energy blending boundaries defined in Fig.~\ref{fig:si-edep}.}
\label{fig:si-loss-relative}
\end{figure}

\begin{figure}[tbp]
\centering
\includegraphics[width=\linewidth]{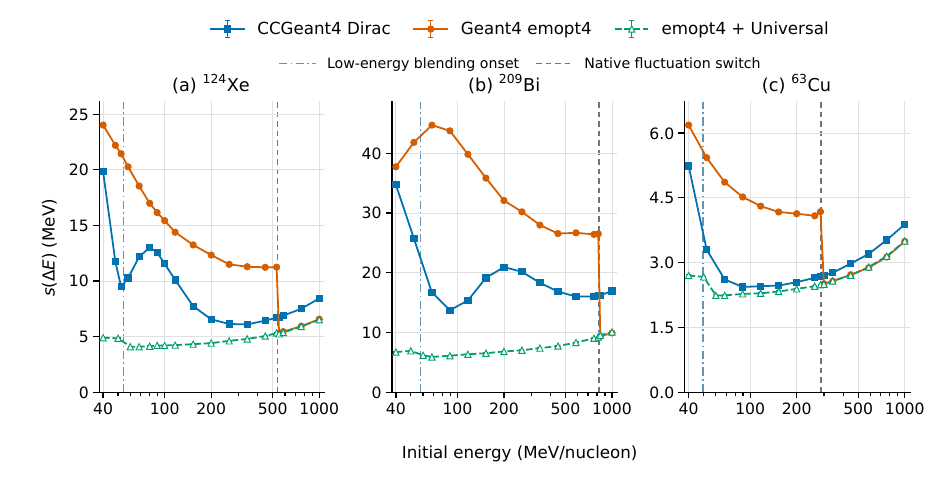}
\caption{Standard deviations $s(\Delta E)$ of the primary-ion energy loss for the samples in Fig.~\ref{fig:si-loss-relative}. The vertical scale differs between panels.}
\label{fig:si-loss-absolute}
\end{figure}

Figures~\ref{fig:si-loss-relative} and~\ref{fig:si-loss-absolute} show the relative and absolute widths of the primary-ion energy loss, $\Delta E=T_{\mathrm{in}}-T_{\mathrm{out}}$. We add a control in which only the \texttt{emopt4} ionization fluctuation model is replaced by \texttt{G4UniversalFluctuation}, retaining the native stopping power and other primary-ion processes.

Replacing the native fluctuation model by Universal removes the abrupt changes, but does not reproduce the CCGeant4 energy dependence for xenon and bismuth. Charge-changing histories broaden the distributions at lower energies, while relativistic Coulomb (Mott) corrections enhance collisional straggling at higher energies, as observed in energy-loss measurements at 700--1000 MeV/u~\cite{Scheidenberger1996}. Delta-electron escape further affects the deposited-energy widths.

\subsection{Dependence on the production cut}
\label{sec:cut-dependence}
\begin{figure}[tbp]
\centering
\includegraphics[width=\linewidth]{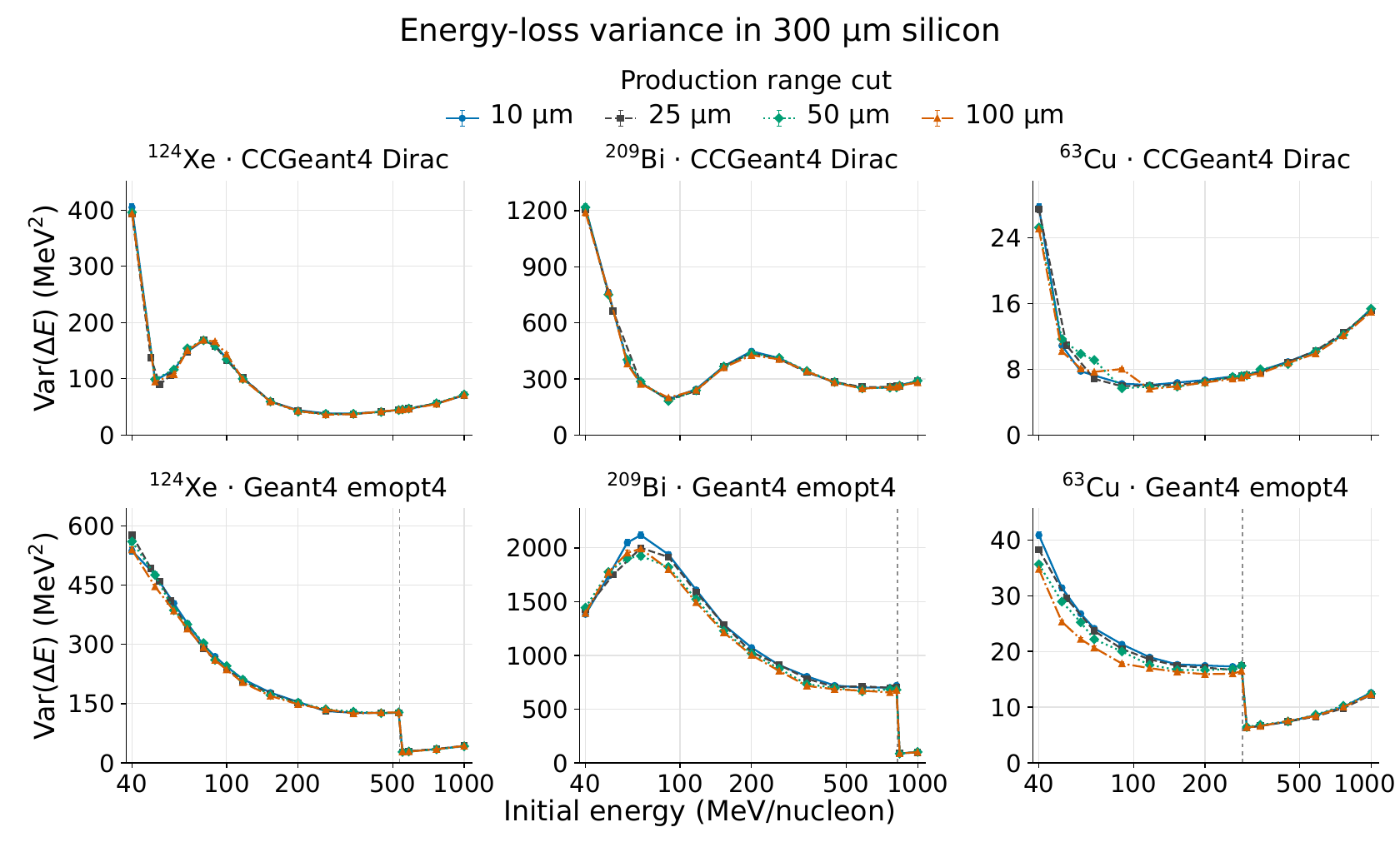}
\caption{Energy-loss variances in 300~$\mu$m of silicon at four production range cuts. The upper row shows CCGeant4 and the lower row standard \texttt{emopt4}. Columns correspond to xenon, bismuth and copper. Grey dashed lines mark the native ion-fluctuation thresholds.}
\label{fig:si-cut-variance}
\end{figure}

Figure~\ref{fig:si-cut-variance} compares the energy-loss variances at production range cuts of 10, 25, 50 and 100~$\mu$m, using the beams and silicon geometry of Section~\ref{sec:standard-em-comparison} and 10\,000 events per point. This cut scan covers incident energies from 40 to 1000~MeV/u. For Xe and Bi in CCGeant4, the differences from the 25~$\mu$m results at matching energies remain within about 8\%. Copper is more sensitive at the lowest energies. The jumps in the copper curves at cuts of 50 and 100~$\mu$m arise from an internal change of regime in \texttt{G4UniversalFluctuation}. Standard \texttt{emopt4} also shows cut dependence.

The cut dependence reflects the division of energy transfers between explicit delta-electron production and continuous fluctuations described in Section~\ref{sec:transitions-fluctuations}.

\subsection{Energy loss and detector response}
\label{sec:detector-response}
\begin{figure}[!htbp]
\centering
\includegraphics[width=\linewidth]{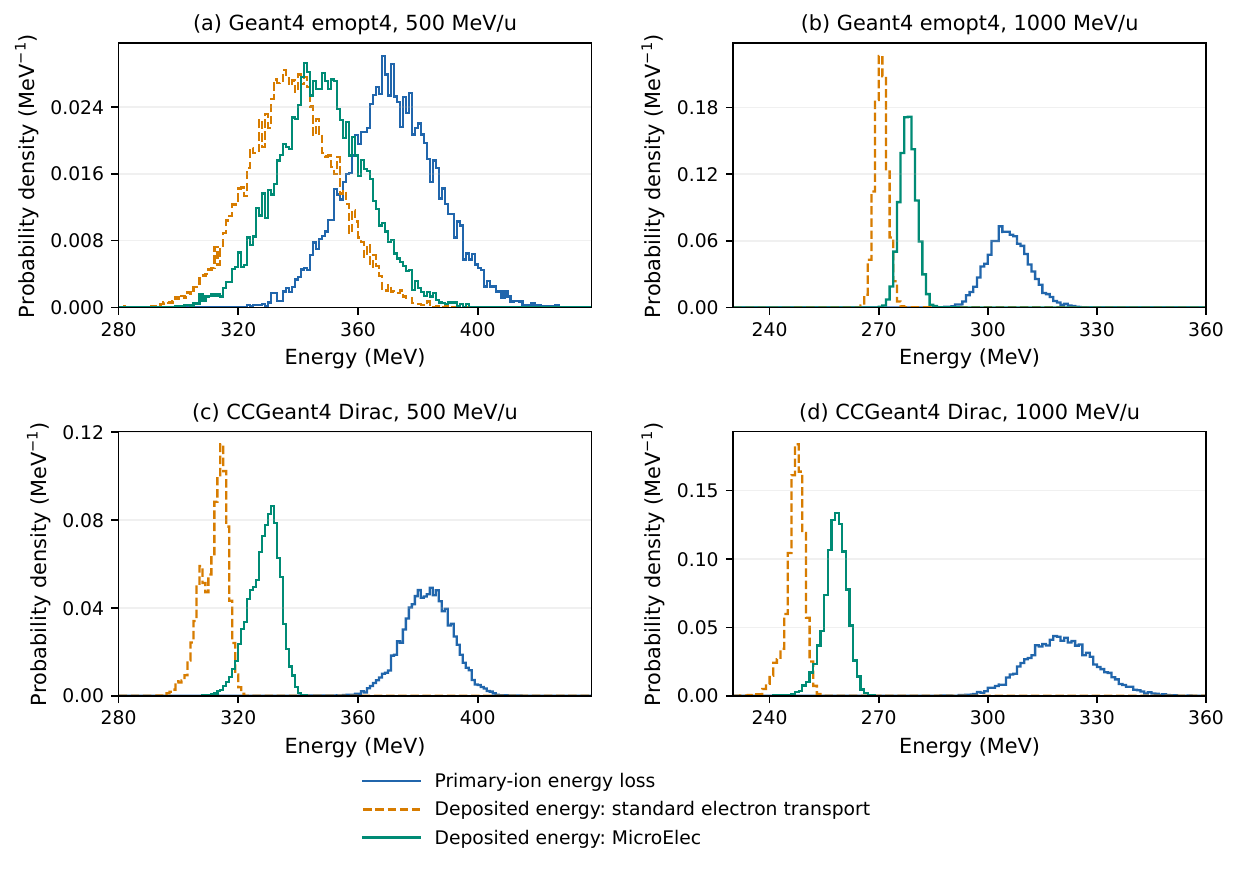}
\caption{Primary-ion energy loss and deposited energy for $^{209}$Bi in $100~\mu$m of silicon. The upper row shows standard \texttt{emopt4}, and the lower row CCGeant4 Dirac, at 500~MeV/u (left) and 1000~MeV/u (right). Deposited-energy spectra are shown for standard electron transport and the MicroElec configuration described in the text. Each distribution is normalized to unit area.}
\label{fig:bi-secondary-transport}
\medskip
\begingroup
\makeatletter\def\@captype{table}\makeatother
\small
\caption{Standard deviations (MeV) of the distributions in Fig.~\ref{fig:bi-secondary-transport}, using the angular selection of Section~\ref{sec:edep-energy-scan}.}
\label{tab:bi-secondary-transport}
\begin{tabular}{rlrrr}
\toprule
$K/A$ (MeV/u) & Model & $\sigma(\Delta E)$ & \multicolumn{2}{c}{$\sigma(E_{\rm dep})$} \\
 & & & Standard transport & MicroElec \\
\midrule
500 & Geant4 \texttt{emopt4} & 15.47 & 14.89 & 14.96 \\
500 & CCGeant4 Dirac & 8.20 & 4.41 & 5.05 \\
1000 & Geant4 \texttt{emopt4} & 5.72 & 1.71 & 2.25 \\
1000 & CCGeant4 Dirac & 9.56 & 2.62 & 3.21 \\
\bottomrule
\end{tabular}
\endgroup
\end{figure}

Figure~\ref{fig:bi-secondary-transport} compares CCGeant4 Dirac and standard \texttt{emopt4} for $^{209}$Bi at 500 and 1000~MeV/u in a $32\times32$~mm$^2$ silicon plate, $100~\mu$m thick. For each model we compare standard electron transport with a MicroElec configuration. The beams enter normally at the centre of the detector, with a 10~$\mu$m production cut and 10\,000 events per sample. In the MicroElec configuration, electrons below the corresponding 31.7~keV production threshold deposit their remaining energy locally. Above 9~MeV, standard electron models are used.

Changing the secondary-electron transport leaves the primary-ion energy-loss distribution unchanged, but shifts and broadens the deposited-energy peak. At 500~MeV/u, \texttt{emopt4} gives broad energy-loss and deposited-energy spectra, whereas the Dirac deposited-energy spectrum is much narrower than its energy-loss spectrum. At 1000~MeV/u, where \texttt{emopt4} has switched to Universal fluctuations, deposited-energy spectra are substantially narrower than energy-loss spectra in both models. The widths are listed in Table~\ref{tab:bi-secondary-transport}.

\section{Discussion and conclusions}
Dirac describes the effect of projectile-electron screening on both stopping and delta-electron spectra. Legacy provides a useful comparison with traditional charge-scaled stopping calculations, including MOCADI. Both retain discrete charge exchange. The electron production cut should be low enough that Dirac corrections to the $q^2$ approximation below the threshold can be neglected. The xenon and bismuth energy-loss variances show weak cut dependence over the tested range (Section~\ref{sec:cut-dependence}). We recommend explicit charge-state transport above about 100 MeV/u within the range covered by the input tables.

The charge-evolution and MOCADI benchmarks test the implementation against the reference descriptions. For uranium at 800 MeV/u, Dirac follows the screened-ion calculation of Weick et al., and both calculations lie within the experimental uncertainties in Be, Al and Pb. For bismuth, both modes reproduce the reversal of the energy-loss asymmetry in Be. The calculated widths agree reasonably with the measurements in several targets given the experimental uncertainties, but clear discrepancies remain in Al at 600 MeV/u (Fig.~\ref{fig:weick-bi}) and in the silver spectra, where both configurations give narrower distributions than measured (Table~\ref{tab:weick-spectra}).

For xenon, Dirac and Legacy give similar central-peak widths, compatible with the experimental upper estimates in silicon. Standard \texttt{emopt4} gives broader peaks than measured. Residual beam and detector broadening prevents these data from determining the quantitative accuracy of the intrinsic energy-deposition widths.

The calculations show substantial differences between energy-loss and deposited-energy distributions in thin silicon, in both width and shape. Charge exchange affects continuous stopping as well as delta-electron production, while electron escape can make the deposited-energy distribution much narrower. A single broad energy-loss peak can also coexist with two deposited-energy maxima (\ref{app:legacy-deposition}). Section~\ref{sec:detector-response} shows that changing secondary-electron transport shifts and broadens the deposited-energy peak while leaving the primary-ion energy-loss distribution essentially unchanged. Quantitative detector-response predictions therefore depend on the secondary-electron transport model, which can be selected or replaced independently in CCGeant4. The charge-changing cross sections, stopping powers and delta-electron spectra can likewise be varied through the input tables.

\section*{Code and data availability}
The code is available on GitHub as \href{https://github.com/andreiazhakin/CCGeant4/releases/tag/v0.1.0-rc1}{CCGeant4 version 0.1.0-rc1}, commit \href{https://github.com/andreiazhakin/CCGeant4/tree/390fb82114d1fc51f62f5bcd2a44d14a910e5a0d}{\texttt{390fb82}}~\cite{CCGeant4Software}. The \href{https://github.com/andreiazhakin/CCGeant4/releases/download/v0.1.0-rc1/CCGeant4-source-20261114.zip}{source package} contains the patch for Geant4~11.4.0, the ionization and elastic-table converters, and a Cu, Xe and U transport example with its input tables. The \href{https://github.com/andreiazhakin/CCGeant4/releases/tag/article-data-20261109}{article data archive} contains simulation events, run configurations and numerical comparison data for the figures and tables. The xenon measurements are described in the companion study~\cite{AzhakinXeDraft}.

The \href{https://github.com/andreiazhakin/CCGeant4/releases/tag/runtime-20261107}{\texttt{runtime-20261107} release} includes five table archives. One archive contains the ionization, stopping and charge-exchange tables. The other four contain the elastic-scattering tables for silicon. The complete runtime tables occupy approximately 11.2 GB on disk, or 5.8 GB in the compressed archives.

After unpacking, the elastic-scattering tables occupy 6.25 GB, the GLOBAL cross sections 3.14 GB and the Dirac electron spectra 1.59 GB. The main ATIMA-normalized stopping tables occupy 203 MB, and the low-energy ATIMA table occupies 52 MB. Most of the disk space is therefore used by the elastic and GLOBAL datasets. The size of the elastic dataset reflects its dense energy and angular grids, used to resolve the Coulomb--nuclear interference pattern when interpolating between tabulated nodes.

The source package includes SHA-256 checksums and source credits. The upstream solvers and coefficient collections are distributed by their respective providers under their own terms.

\section*{Author contributions and declarations}
\textbf{CRediT author contributions.} Andrei Azhakin: Software, Investigation, Validation, Visualization, Writing---original draft. Ilya Kudryashov: Supervision.

\textbf{Funding.} The study was conducted under the state assignment of Lomonosov Moscow State University.

\textbf{Declaration of competing interests.} The authors declare that they have no known competing financial interests or personal relationships that could have appeared to influence the work reported in this paper.

\appendix
\counterwithin*{figure}{section}
\counterwithin*{table}{section}
\section{Mean energy loss above the Dirac tables}
\label{app:ls-normalization}
Table~\ref{tab:ls-normalization} compares standard Geant4 and the CCGeant4 high-energy fallback with ATIMA~1.41 for a bare $^{209}$Bi ion traversing $100~\mu$m of silicon at normal incidence. The observable is the mean kinetic energy lost by the primary ion, $\langle T_{\rm in}-T_{\rm out}\rangle$. All ions remain above the upper transition at $\Eu=2000$~MeV throughout the layer. For ATIMA, we obtain the exit energy from the range--energy relation for Bi in Si at the same incident speed and areal density~\cite{AtimaWeb}.

\begin{table}[!htbp]
\centering
\caption{Mean energy loss (MeV per ion) in $100~\mu$m of silicon. Each transport entry is the mean and its statistical error for 10\,000 ions.}
\label{tab:ls-normalization}
\small
\begin{tabular}{rrrr}
\toprule
$K/A$ (GeV/u) & Geant4 11.4.0 \texttt{emopt4} & ATIMA 1.41 & CCGeant4 fallback \\
\midrule
2.5 & $284.1\pm0.1$ & 299.0 & $299.5\pm0.2$ \\
4.0 & $288.2\pm0.1$ & 303.2 & $304.0\pm0.2$ \\
6.0 & $295.8\pm0.2$ & 310.8 & $312.0\pm0.3$ \\
8.0 & $303.2\pm0.2$ & 317.3 & $319.0\pm0.4$ \\
\bottomrule
\end{tabular}
\end{table}

Across these four energies, the CCGeant4 means agree with ATIMA to within 0.6\% and the standard Geant4 means are 4.4--5.0\% below it. The models treat the correction to first-Born stopping differently. CCGeant4 calculates it using a Mott electron spectrum, the Bloch correction and a finite-nuclear-size correction. Standard Geant4 uses a tabulated Lindhard--S\o rensen term. One difference is its normalization. In the conventional bare-ion stopping formula the coefficient of $\Delta L_{\rm LS}$ is $4\pi r_e^2 m_e c^2 n_e Z_{\rm p}^2/\beta^2$. The Geant4 11.4.0 and 11.4.3 implementations use half this coefficient for the tabulated term~\cite{Geant4LSSource,Kats2021}. For example, at $Z_{\rm p}=36$ and $\beta=0.95$ the table returns 0.441, close to the published point-nucleus value of 0.443~\cite{Kats2021}. The table thus uses the conventional normalization, so the coefficient multiplying it in these Geant4 releases is too small by a factor of two.

\section{Numerical conventions and data preparation}
\label{app:numerics}
The converter combines the original Dirac spectra, GLOBAL capture and loss cross sections, and ATIMA 1.41 range curves. The Dirac ionization dataset covers $Z_{\mathrm p}=29$--92, $Z_t=1$--84 and 88--96, $\Eu=30$--2000 MeV and $N=0$--27. The supplied GLOBAL and Legacy stopping tables extend the projectile range to $Z_{\mathrm p}=96$ with the same target elements. Charge states beyond the directly calculated subset are treated as described in Section~\ref{sec:charge-continuation}. The elastic reference tables cover $Z_{\mathrm p}=26$--82 on silicon at $K/A=10$--5000 MeV/u. The interface also accepts tables for other projectile--target pairs.

ATIMA gives the range $R$ in mg/cm$^2$ as a function of kinetic energy per atomic-mass unit, $E$ in MeV/u. The converter calculates the stopping power from the slope of this range curve:
\begin{equation}
 S_m(E)=\frac{1000\,M_{\mathrm{ref}}}{dR/dE},\qquad
 \epsilon_t(E)=\frac{A_{t,\mathrm{ref}}}{N_{\mathrm A}}S_m(E).
\end{equation}
Here $M_{\mathrm{ref}}$ is the projectile mass used by ATIMA to calculate that range curve, in atomic-mass units. The target molar mass used by ATIMA is $A_{t,\mathrm{ref}}$ in g/mol, and $N_{\mathrm A}$ is the Avogadro constant. Thus $S_m$ is the stopping power per unit areal density, in MeV\,cm$^2$/g, and $\epsilon_t$ is the stopping cross section per target atom, in MeV\,cm$^2$.

The resulting stopping values are interpolated between tabulated energies. The low-energy ATIMA tables use the same masses and energy convention. Above 30 MeV/u, ATIMA uses Lindhard--S\o rensen stopping with additional atomic corrections~\cite{AtimaWeb,LindhardSorensen1996,ScheidenbergerGeissel1998}.

Between tabulated electron energies, the Dirac spectrum is described by a positive power law. Its integrals give the collision probabilities and energy-loss moments, and random electron energies are drawn from its cumulative distribution. The spectra are available at 39 incident energies from 30 to 2000 MeV. The stopping tables use a separately refined energy grid.

At an incident energy between two tabulated values, we combine the two neighbouring spectra with positive weights that vary linearly with the logarithm of incident energy. First, their electron-energy intervals are rescaled to the kinematic limit at the requested energy, keeping the same relative position $u=(T-T_0)/(\Tmax-T_0)$. The kinematic limit uses the actual projectile mass, so this rescaling also accounts for its effect on the maximum electron energy. The spectrum height is adjusted so that this change of interval preserves the integrated cross section. To generate an electron above the production threshold, the two spectra are selected in proportion to their interpolation weights multiplied by their cross sections above that threshold.

For the same tabulated inputs, the spectral moments and stopping powers from the C++ implementation agreed with an independent Python calculation to a relative difference below $10^{-12}$ at the tested points. In 132 sampling tests for Cu, Xe and U, the mean energy of the sampled delta electrons differed from the mean obtained by integrating the tabulated spectrum by at most 3.34 standard errors.

For Cu, Xe and U, we omitted each interior energy node in turn and reconstructed the spectrum from its two neighbours. At the four electron thresholds used in Section~\ref{sec:cut-dependence}, the above-threshold cross section and its first two energy moments differed from the omitted values by at most 4.5\%, or 3.2\% for $\Eu\ge100$ MeV. The energy grid can be refined by calculating additional Dirac spectra.

The soft-stopping coefficient $B$ is interpolated linearly in logarithmic incident energy. GLOBAL cross sections are interpolated in logarithmic energy as well: we interpolate their logarithms when both values are positive, and the cross sections themselves when either value is zero.

Where the Dirac and fallback models overlap, the continuous stopping is
\begin{equation}
 S_{\mathrm{restricted}}=w(S_{\mathrm D}-H_{\mathrm D})+(1-w)(S_{\mathrm F}-H_{\mathrm F}).
\end{equation}
Here D and F refer to the Dirac and fallback models, $S$ is the full electronic stopping, and $H$ is the energy transferred to explicitly generated electrons per unit path. These $H$ terms use the production threshold. With the default settings, the same weights govern the probabilistic choice of the continuous-fluctuation model (Section~\ref{sec:transitions-fluctuations}).

A production threshold below 0.99 keV is rejected because it lies outside the supported Dirac spectrum. This check uses the electron energy obtained from Geant4's range cut for each material and set of cuts.

\section{Construction and use of the elastic reference tables}
\label{app:elastic}
The supplied tables for 57 projectile--silicon pairs are an example for the tabulated scattering interface. At 10--50 MeV/u, we use the REGINA SPP2 optical potential~\cite{Chamon2021,ReginaData,Ribeiro2026}. From 50 to 400 MeV/u, we use MGOP~\cite{Furumoto2012}, based on the CEG07b effective interaction~\cite{Furumoto2008,Furumoto2009}. Both optical-potential calculations are solved with FRESCOX~\cite{Thompson1988}. Above 400 MeV/u, we use an optical-limit Glauber calculation~\cite{Mehndiratta2017}, with SAID SM16 nucleon--nucleon input~\cite{Workman2016} up to 1300 MeV/u and the Uzhinsky parametrization~\cite{Uzhinsky2016} at higher energies.
These sources can also be used to prepare tables for other projectile--target pairs within their ranges of applicability. Details of the calculations and the smooth joining of energy ranges are documented with the dataset.

At each energy and angle, the input is the ratio $R$ of the calculated differential cross section to the Rutherford cross section. It includes Coulomb--nuclear interference but not atomic screening. Screening is then included through
\begin{equation}
 \frac{d\sigma}{dx}=\frac{C\,R(x)}{(A_s+x)^2},\qquad x=\sin^2(\theta_{\mathrm{CM}}/2),
\end{equation}
where $A_s$ is the Wentzel--Moli\`ere screening parameter and $C$ sets the Rutherford normalization.

The tables store $R$ together with the total cross section and cumulative angular probabilities obtained from this screened distribution. The transport model uses them to determine collision rates and generate scattering angles, with linear interpolation of $R$ in $x$. Between tabulated energies, it combines the neighbouring differential cross sections with weights linear in kinetic energy per nucleon.

\begin{figure}[!htbp]
\centering
\includegraphics[width=\linewidth]{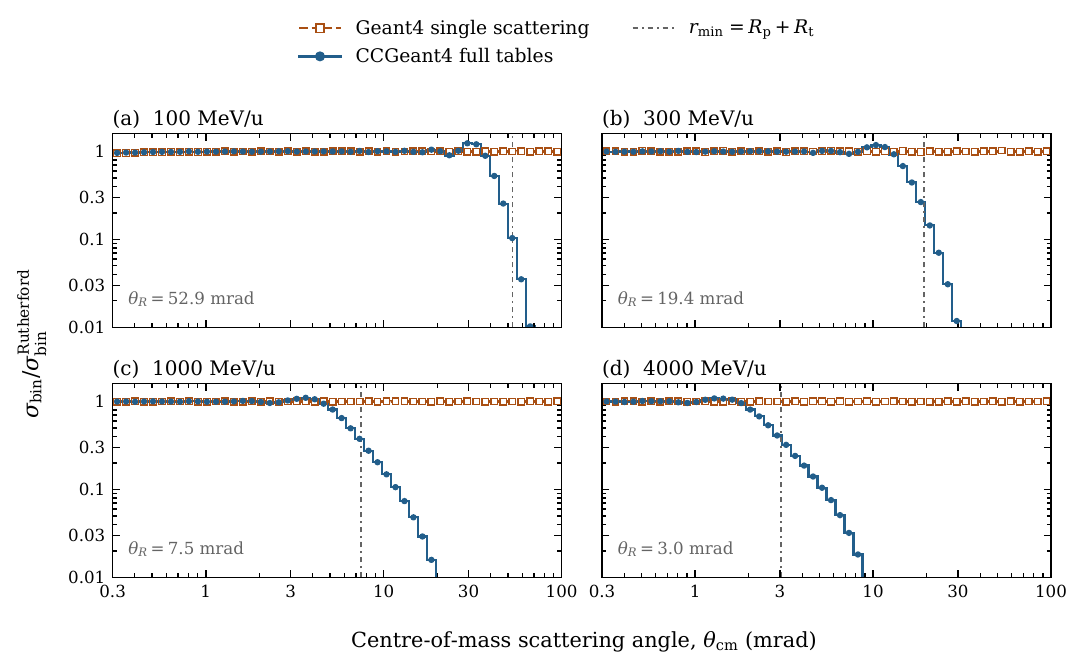}
\caption{Single elastic scattering of $^{132}\mathrm{Xe}$ on $^{28}\mathrm{Si}$ at 100, 300, 1000 and 4000~MeV/u. Lines show the cross sections from native Geant4 ion Coulomb scattering and the full CCGeant4 silicon tables, each divided by the unscreened Rutherford cross section in the same angular bin. Points show Monte Carlo estimates with one standard error. Vertical grey dash-dotted lines mark $r_{\min}=R_{\mathrm p}+R_t$ for the reference Coulomb orbit.}
\label{fig:xe-si-single-scattering}
\end{figure}

The vertical lines in Fig.~\ref{fig:xe-si-single-scattering} mark the Coulomb scattering angle $\theta_R$ for a distance of closest approach equal to the sum of the nuclear radii, $R_{\mathrm c}=1.2(A_{\mathrm p}^{1/3}+A_t^{1/3})$~fm. For a Rutherford trajectory,
\begin{equation}
 r_{\min}=a\left[1+\frac{1}{\sin(\theta_{\mathrm{CM}}/2)}\right],\qquad
 \theta_R=2\arcsin\!\left(\frac{a}{R_{\mathrm c}-a}\right).
\end{equation}
Here $a=\sqrt{C/\pi}$ is the Coulomb length determined by the Rutherford normalization. As a rough geometric approximation, one can suppress the screened Coulomb cross section above $\theta_R$, using a sharp or smooth cutoff instead of resolving the interference pattern.

\section{Additional energy-loss and deposited-energy comparisons}
\label{app:legacy-deposition}
\begin{figure}[!htbp]
\centering
\includegraphics[width=\linewidth]{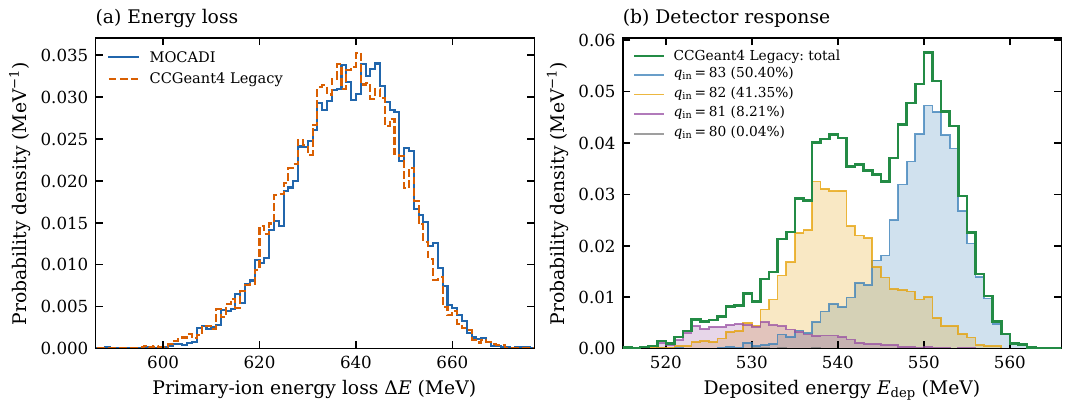}
\caption{Energy loss and energy deposited in $150\,\mu\mathrm{m}$ of silicon
for $^{209}\mathrm{Bi}$ at $380.416\,\mathrm{MeV}$ per nucleon. The
incident equilibrium charge mixture is the same as in the MOCADI benchmark.
The full distributions have unit area. Shaded components in (b) are
grouped by entrance charge.}
\label{fig:legacy-loss-edep}
\end{figure}

Figure~\ref{fig:legacy-loss-edep} uses the CCGeant4 Legacy runs of Section~\ref{sec:legacy-cross-code}.

The difference between these quantities is particularly clear for bismuth
at $380.416\,\mathrm{MeV}$ per nucleon in $150\,\mu\mathrm{m}$ of silicon
(Fig.~\ref{fig:legacy-loss-edep}). The energy-loss spectrum is a single broad peak,
whereas the deposited-energy spectrum has two resolved maxima. Its standard
deviation is $8.14\,\mathrm{MeV}$, compared with $11.68\,\mathrm{MeV}$
for the primary-ion energy loss. Charge-dependent stopping and the escape
of secondary electrons therefore affect the shape as well as the width
of the detector signal. Decomposing the deposited-energy spectrum by
entrance charge shows that the lower-energy maximum is dominated by
$q_{\mathrm{in}}=82$ and the higher-energy maximum by $q_{\mathrm{in}}=83$.

At $1000\,\mathrm{MeV}$ per nucleon in $300\,\mu\mathrm{m}$ of silicon,
the deposited-energy standard deviations are 3.00 and $5.73\,\mathrm{MeV}$
for xenon and bismuth, respectively, about 2.8--2.9 times smaller than the
corresponding energy-loss widths (Table~\ref{tab:legacy-mocadi}).

\section{Bismuth energy-loss variances relative to collisional straggling}
\label{app:bi-normalized}
\begin{figure}[!htbp]
\centering
\includegraphics[width=\linewidth]{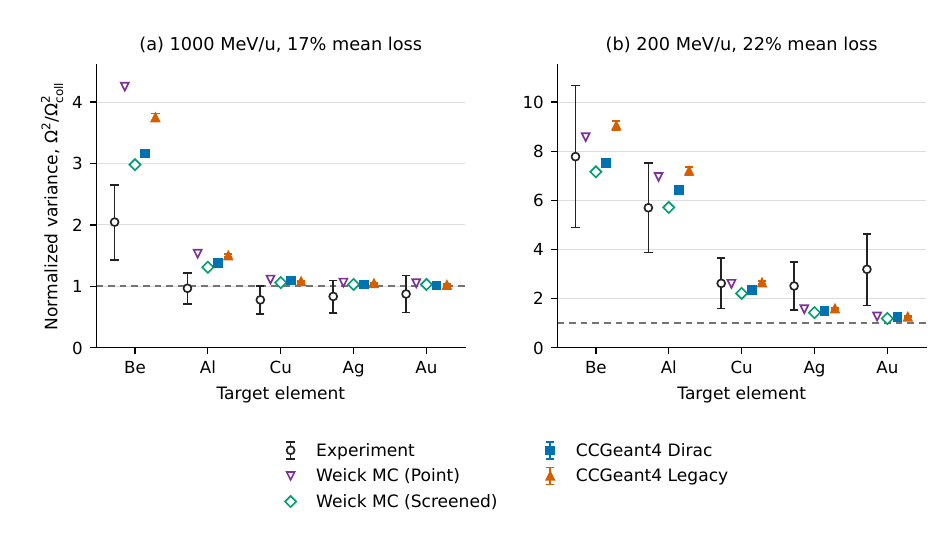}
\caption{Energy-loss variances of $^{209}$Bi in Be, Al, Cu, Ag and Au normalized to the Lindhard--S\o rensen collisional variance. The measurements and their error bars are from Figs.~11 and 12 of Ref.~\cite{WeickNIM2000}. The Point and Screened Monte Carlo results are from Figs.~46 and 47 of Ref.~\cite{GeisselWeick2002}. Unity corresponds to collisional straggling alone. CCGeant4 bars show Monte Carlo statistical uncertainties.}
\label{fig:weick-bi-normalized}
\end{figure}

Figure~\ref{fig:weick-bi-normalized} compares the five targets of Fig.~\ref{fig:weick-bi} at 200 and 1000 MeV/u in the normalization used by Weick et al.~\cite{WeickNIM2000}. The published mean energy losses were 22\% and 17\% of the incident energy, respectively. The two Weick Monte Carlo variants treat the projectile either as a point charge or with a realistic electronic charge distribution~\cite{GeisselWeick2002}.

\section{Computational performance}
\label{sec:performance}
We timed 10\,000 $^{238}$U ions at 100 and 1500~MeV/u in 175~$\mu$m of silicon and 60~$\mu$m of lead on an AMD Ryzen~7~7700 processor, using one or eight worker threads. The ions enter normally, secondary particles are fully transported, and hadronic reactions are disabled. CCGeant4 Dirac uses GLOBAL equilibrium entrance-charge fractions. A production range cut of 25~$\mu$m is compared with a high cut of $10^5~\mu$m, which suppresses explicit delta-electron production. The measured interval starts after initialization and warm-up and ends with the last event, before result merging and output.

\begin{table}[!htbp]
\centering
\caption{Representative transport times in seconds for 10\,000 uranium ions. $E=K/A$ is in MeV/u, and the cut is in $\mu$m. Columns marked 1 and 8 give the numbers of worker threads.}
\label{tab:performance}
\begin{tabular}{lrr rrrr}
\toprule
Target & $E$ & Cut & \multicolumn{2}{c}{Geant4 \texttt{emopt4}} & \multicolumn{2}{c}{CCGeant4 Dirac}\\
\cmidrule(lr){4-5}\cmidrule(lr){6-7}
 & & & 1 & 8 & 1 & 8\\
\midrule
Si & 100  & 25     & 187.4 & 34.9 & 666.7 & 106.1\\
Si & 100  & $10^5$ & 0.161 & 0.027 & 0.712 & 0.131\\
Si & 1500 & 25     & 83.5 & 12.3 & 167.5 & 24.6\\
Si & 1500 & $10^5$ & 0.150 & 0.028 & 0.252 & 0.042\\
Pb & 100  & 25     & 224.4 & 36.6 & 840.2 & 122.8\\
Pb & 100  & $10^5$ & 0.163 & 0.028 & 11.86 & 1.60\\
Pb & 1500 & 25     & 114.9 & 17.0 & 191.9 & 28.6\\
Pb & 1500 & $10^5$ & 0.154 & 0.023 & 0.271 & 0.047\\
\bottomrule
\end{tabular}
\end{table}

At the 25~$\mu$m cut, CCGeant4 requires about 1.7--3.7 times the single-thread transport time of \texttt{emopt4}. Using eight threads accelerates CCGeant4 by a factor of 6.3--6.8 (Table~\ref{tab:performance}). Analytic estimates along the mean slowing trajectory give about 12 and 286 charge-changing collisions per ion at 100~MeV/u in Si and Pb, respectively, falling to 0.16 and 0.74 at 1500~MeV/u. The corresponding numbers of explicit delta electrons at the 25~$\mu$m cut range from about 500 to 2800 per ion. Suppressing their production greatly reduces the transport time. The remaining cost is largest for low-energy uranium in lead, where charge exchange is frequent. Peak resident memory of the whole process was 96--150~MiB for CCGeant4 and 43--66~MiB for \texttt{emopt4}, with only the table subset needed for these benchmarks loaded.

\section*{Declaration of generative AI and AI-assisted technologies in the manuscript preparation process}
During the preparation of this work, the authors used OpenAI Codex (OpenAI) for assistance with code development and review, data analysis and plotting, source review, and drafting and revising the manuscript. The authors reviewed and edited the output as needed and take full responsibility for the content of the published article.

\end{document}